%% file: pr427_ff_journal.tex
\documentclass[11pt]{article}
\usepackage{a4wide,epsfig,rotating,cite,amsmath,amssymb}

\begin{document}

\input{pr427_f_defini}

%%%%%%%%% TITLE

\begin{titlepage}
\begin{center}
{\large EUROPEAN ORGANIZATION FOR NUCLEAR RESEARCH}
\end{center}

\bigskip
\begin{flushright}
CERN-PH-EP/2008-016 \\
14 October 2008 \\
%updated: 9 March 2010 \\
%updated: 21 February 2012 
updated: 21 June 2012 
% finalization for submission: September 15, 2008 

\end{flushright}

\bigskip\bigskip\bigskip
\begin{center}
{\huge\bf\boldmath
      Search for Charged Higgs Bosons \\ 
      in \ee\ Collisions at $\sqrts=189-209$~GeV
\unboldmath}
\end{center}

\bigskip\bigskip
\begin{center}
{\LARGE The OPAL Collaboration}
\end{center}

\bigskip\bigskip\bigskip
\begin{center}
{\large  Abstract}
\end{center}

A search is made for charged Higgs bosons predicted by Two-Higgs-Doublet
extensions of the Standard Model (2HDM) using electron-positron collision data
collected by the OPAL experiment at $\sqrts=189-209$~GeV,  corresponding to
an integrated luminosity of approximately 600~\pb.  Charged Higgs bosons are
assumed to be pair-produced   and to decay into \qq, \tnt\ or \AWpm. No signal
is observed.  
Model-independent limits on the charged Higgs-boson production cross section are
derived by combining these results with previous searches at lower energies.  
%
% Excluded areas on the \mHBR\ plane are presented assuming \BRsum.
% Under the above assumption,  motivated by general 2HDM type II models, charged
%
Under the assumption \BRsum, motivated by general 2HDM type II models,
excluded areas on the \mHBR\ plane are presented and charged
Higgs bosons  are excluded up to a mass of 76.3~GeV at 95\% confidence
level, independent of the branching ratio \BRtn. A scan of the 2HDM type I
model parameter space is performed and limits on the Higgs-boson masses \mHpm\
and \mA\ are presented  for different choices of \tanb.

\bigskip\bigskip\bigskip
\begin{center}
{\bf Submitted to Eur. Phys. J. C}

\end{center}
\end{titlepage}

\input{pr427_f_author}

%%%%%%%%% INTRO

\section{Introduction}\label{sect:intro}

In the Standard Model (SM)~\cite{sm}, the electroweak symmetry  is broken
via the Higgs mechanism~\cite{higgs} generating the masses of elementary
particles.  This requires the introduction of a complex scalar Higgs-field
doublet and implies the existence of a single neutral scalar particle, the
Higgs boson. While the \SM\ accurately describes the  interactions between
elementary particles, it leaves several fundamental questions unanswered. 
Therefore, it is of great interest to study extended models.

The minimal extension of the \SM\ Higgs sector required, for example, by
supersymmetric models contains two Higgs-field doublets~\cite{higgshunter}
resulting in five Higgs bosons: two charged (\Hpm) and three neutral. If
CP-conservation is assumed, the three neutral Higgs bosons are CP-eigenstates:
h and H are CP-even and A is CP-odd. Two-Higgs-Doublet Models (2HDMs) are
classified according to the Higgs-fermion coupling structure. In type I
models (\THDMI)~\cite{THDMI}, 
all quarks and leptons couple to the same Higgs doublet, while
in type II models (\THDMII)~\cite{THDMII}, 
down-type fermions couple to the first Higgs doublet, 
and up-type fermions to the second. 

Charged Higgs bosons are expected to be pair-produced in the process 
\ee\ra\HH\ at LEP, the reaction \ee\ra\Hpm W$^{\mp}$ having a much lower
cross section \cite{kanemura}. In 2HDMs, the tree-level
cross section~\cite{hpmxsec} for pair production is completely determined by
the charged Higgs-boson mass and known \SM\ parameters. 

The \Hpm\ branching ratios are model-dependent. In most of the \THDMII\
parameter space, charged Higgs bosons decay into the heaviest kinematically
allowed fermions, namely \tnt\ and quark pairs\footnote{ 
Throughout this paper charge conjugation is implied. For simplicity, the
notation \tnt\ stands for $\tau^+\nu_\tau$ and  $\tau^-\bar{\nu}_\tau$ and \qq\
for a quark and anti-quark of any flavor combination.}.  
The situation changes in \THDMI, where the decay  \mbox{\Hpm\ra A\Wpms}
can become dominant if the ratio of the vacuum expectation values of the two
Higgs-field doublets is such that \tanb$\gtrapprox$1 and the A boson is
sufficiently light~\cite{THDM-Akeroyd}.

In this paper we search for charged Higgs bosons decaying into \qq, \tnt\
and A\Wpms\ using the data collected by the OPAL Collaboration in 1998$-$2000.  
The results are interpreted within general \THDMII\ assuming 
\BRsum\ for the  branching ratios and in \THDMI\ taking into account
decays  of charged Higgs bosons via A\Wpms, as well. Our result is not confined to 
\qq=\{c\=s, \=cs\}  although that is the dominant hadronic decay channel in most of
the parameter space.

The previously published OPAL lower limit on the charged Higgs-boson mass,
under the assumption of \BRsum, is $\mHpm>59.5$~GeV at 95\% confidence
level (\CL) using data collected at $\sqrts\leq 183$~GeV~\cite{hpmpaper172,hpaper183}.
Lower bounds of $74.4-79.3$~GeV have been
reported by the other LEP collaborations~\cite{hpmaleph, hpmdelphi, hpml3} 
based on the full LEP2 data set. The DELPHI Collaboration also performed a search
for \Hpm\ra\AWpm\ decay and constrained the charged Higgs-boson mass in 
\THDMI~\cite{hpmdelphi} to be \mHpm$\geq$76.7~GeV at 95\%\CL.

%%%%%%%%% EXPERIM

\section{Experimental considerations} \label{sect:detector}

The OPAL detector
is described in~\cite{detector}. 
The events are reconstructed from charged-particle tracks and
energy deposits (\textsl{clusters}) in the electromagnetic and hadron calorimeters.
The tracks and clusters must pass a set of quality requirements
similar to those used in previous OPAL Higgs-boson searches~\cite{SMhiggs}.
In calculating the total visible energies and momenta of events and
individual jets, corrections are applied to prevent 
double-counting of energy in the case of tracks and associated
clusters~\cite{SMhiggs}. 

The data analyzed in this paper were collected in 1998$-$2000 at
center-of-mass energies of $189-209$~GeV as given in Table~\ref{tab:lumi}.  Due
to different requirements on the operational state of the OPAL subdetectors,
the integrated luminosity of about 600~\pb\ differs slightly among
search channels. 

\begin{table}[htbp!]
\begin{center}
{\footnotesize
\begin{tabular}{|l||c|c|c|c|c|c|}
\hline
Year & 1998 & \multicolumn{4}{c|}{1999} & 2000 \\
\hline
$E_\mathrm{cm}$ (GeV) & 186$-$190 & 190$-$194 & 194$-$198 & 198$-$201 
& 201$-$203 & 200$-$209\\
\hline\hline
$<\!E_\mathrm{cm}\!>$ (GeV) & 188.6 & 191.6 & 195.5 & 199.5 & 201.9 & 206.0  \\
\hline
$E_\mathrm{cm}^\mathrm{MC}$ (GeV) & 189 & 192 & 196 & 200 & 202 & 206 \\
\hline
$\int{\!\!\LH}dt$ (\pb) (\twot) & 183.5 & 29.3 & 76.4 & 76.6 & 45.5 & 212.6 \\
\hline
$\int{\!\!\LH}dt$ (\pb) (\twojt, \fourj) & 179.6 & 29.3 & 76.3 & 75.9 & 36.6 & 217.4 \\
\hline
$\int{\!\!\LH}dt$ (\pb) (\eightj, \sixjl, \fourjt) & 175.0 & 28.9 & 74.8 & 77.2 & 36.1 & 211.1 \\
\hline
\end{tabular}
}
\end{center}
\caption{\sl 
Data-taking year, center-of-mass energy bins, 
luminosity-weighted average center-of-mass energies, 
the energies of signal and background Monte Carlo simulations,
and integrated luminosities of the data. 
The data correspond to total integrated 
luminosities of 623.9~\pb\ for the two-tau, 
615.1~\pb\ for the two-jet plus tau and the 
four-jet channels and 
603.1~\pb\ for the \mbox{\Hpm\ra A\Wpms} selections.
}
\label{tab:lumi}
\end{table}

In this paper the following
final states are sought: 
\begin{list}
{$\bullet$}{\itemsep=0pt \parsep=0pt \topsep=0pt}
\item \HH \ra \tpnu\tmnu\ 
    (\textsl{two-tau final state, \twot}), 
\item \HH \ra \qq\tnt\ 
    (\textsl{two-jet plus tau final state, \twojt}), 
\item \HH \ra \qq\qq\  
    (\textsl{four-jet final state, \fourj}),
\item \HH \ra \AWp\AWm \ra \bb\qq\bb\qq\ 
    (\textsl{eight-jet final state, \eightj}), 
\item \HH \ra \AWp\AWm \ra \bb\qq\bb\lnl\ 
    (\textsl{six-jet plus lepton final state, \sixjl}),
\item \HH \linebreak[0] \ra \AWpm\tnt \ra \bb\qq\tnt\ 
    (\textsl{four-jet plus tau final state, \fourjt}). 
\end{list}    

The signal detection efficiencies and accepted background
cross sections are estimated using a variety of Monte Carlo samples.
The HZHA
generator~\cite{hzha} is used to simulate \HH\ production
at fixed values of the charged Higgs-boson mass 
in steps of $1-5$~GeV from the kinematic limit down to
50~GeV for fermionic decays and 40 GeV for bosonic decays.

The background processes are simulated primarily by
the following event generators:
PYTHIA~\cite{pythia} and KK2F~\cite{kk2f} (Z/$\gamma^*$\ra\qq($\gamma$)), 
grc4f~\cite{grc4f} (four-fermion processes, 4f),
BHWIDE~\cite{bhwide} and TEEGG~\cite{teegg} (\ee$(\gamma)$),
KORALZ~\cite{koralz} and KK2F  ($\mu^+\mu^-(\gamma)$ and $\tau^+\tau^-(\gamma)$),
PHOJET~\cite{phojet}, HERWIG~\cite{herwig},
Vermaseren~\cite{vermaseren}
(hadronic and leptonic two-photon processes). 

The generated partons, both for the signal and the \SM\ Monte Carlo
simulations, are hadronized using JETSET~\cite{pythia},  with parameters
described in~\cite{opaltune}. For systematic studies, cluster fragmentation
implemented in HERWIG for the process Z/$\gamma^*$\ra\qq($\gamma$) is used. 
The predictions of 4f processes are cross-checked using 
EXCALIBUR~\cite{excalibur}, KoralW~\cite{koralw} and KandY~\cite{KandY}. 

The obtained Monte Carlo samples are processed through a full simulation of the OPAL
detector~\cite{gopal}. The event selection is described below.

%%%%%%%%% FERMIONIC

\section{Search for four-fermion final states}\label{sect:fermionic}

In most of the parameter space of \THDMII\ and with a sufficiently  heavy A
boson in \THDMI, the fermionic decays of the charged Higgs boson dominate
and lead to four-fermion final states. The most important decay mode  is
typically \Hpm\ra\tnt, with the hadronic mode \Hpm\ra\qq\  reaching about 40\%
branching ratio at maximum. 

The search for the fully leptonic final state \HH\ra\tpnu\tmnu\ is described
in~\cite{acoplanar2003}.  The searches for the \HH\ra\qq\tnt\ and the
\HH\ra\qq\qq\ events are optimized using Monte Carlo simulation of \Hp\ra\cs\
decays. The sensitivities to other quark flavors are similar and the possible
differences are taken into account as systematic uncertainties. 
Therefore, our results are valid for any hadronic decay of the charged Higgs 
boson. 

Four-fermion final states originating from \HH\ production would
have very similar kinematic properties to \WW\ production, which
therefore constitutes an irreducible background to our searches, 
especially when \mHpm\ is close to \mWpm.
To suppress this difficult \SM\ background, a mass-dependent likelihood 
selection
(similar to the technique described in \cite{2hdm00}) is introduced. For each
charged Higgs-boson mass tested (\mtest), 
a specific analysis optimized for a reference
mass (\mref) close to the hypothesized value is used.

We have chosen a set of reference charged Higgs-boson masses at which 
signal samples are generated. Around these reference points, mass regions
(labeled by \mref) are
defined with the borders centered between the neighboring points.  For each
individual mass region, at each center-of-mass energy, we create a separate
likelihood selection. 
The definition of the likelihood function is based on a
set of histograms of channel specific observables, given in~\cite{hpaper183}.
The signal histograms are built using events generated at \mref.  
The background histograms are
composed of the \SM\ processes and are identical for all mass regions.

When testing the hypothesis of a signal with mass \mtest, the background and
data rate and discriminant (i.e. the reconstructed Higgs-boson mass) 
distribution depend on the mass region to which \mtest\ belongs. 
The signal quantities depend on the 
value of \mtest\ itself and are determined as follows. 
The signal rate and discriminant distribution are computed, with the
likelihood selection optimized for \mref, for three
simulated signal samples with masses \mlow, \mref\ and  \mhigh. Here, 
\mlow\ and \mhigh\ are the closest
mass points to  \mref\ at which signal Monte Carlo samples are generated, with
\mlow $<$ \mref $<$ \mhigh. The signal rate and discriminant distribution for
\mtest\ are then calculated by linear interpolation from the quantities for
\mlow\ and \mref\ if \mtest $<$ \mref, or for \mref\ and \mhigh\ if \mtest $>$
\mref.

When building the likelihood function three event classes are considered: 
signal, four-fermion background (including two-photon processes)
and two-fermion background. The likelihood
output gives the probability that a given event belongs to the signal
rather than to one of the two background sources. 

%%%%%%%%% 2JETTAU

\subsection{The two-jet plus tau final state}\label{sect:2jet-tau}

The analysis closely follows our published one at
$\sqrts=183$~GeV~\cite{hpaper183}. It proceeds in two steps.   First, events
consistent with the final state topology of an isolated tau lepton, a pair of
hadronic jets and sizable missing energy are preselected and are then 
processed by a likelihood selection. The sensitivity of the likelihood selection
is improved by
building mass-dependent discriminant functions as explained above. 

Events are selected if their likelihood output (\LH) is greater than a
cut value chosen to maximize the sensitivity of the selection at each simulated
charged Higgs-boson mass (\mref). 
Apart from the neighborhood of the \WW\ peak, the optimal cut
does not depend significantly on the simulated mass 
and is chosen to be \LH $>$ 0.85. 
Around the \WW\ peak, it is gradually reduced to 0.6.

The number of selected events per year is given in Table~\ref{tab:events}
for a test mass of \mHpm=75~GeV.
In total, 
% At the end of the selection, 
331 events are selected in the data sample with
$316.9 \pm 3.2$~(stat.)~$\pm 38.4$~(syst.) events expected from \SM\
processes.
% for a test mass of \mHpm=75~GeV. 
The sources of systematic uncertainties are discussed below. 
Four-fermion processes account for
more than 99\% of the \SM\ background 
and result in a large peak in the reconstructed mass centered at
the \Wpm\ mass (with a second peak at the Z mass for test masses of 
$\mHpm>85$~GeV).  The
signal detection efficiencies for the various  LEP energies  are between 25\%
and 53\% for any charged Higgs-boson mass. 

\begin{table}[hb!]
\begin{center}
\begin{tabular}{|l||l|l||l|l|}\hline
LEP energy           & \multicolumn{2}{l||}{\twojt} & \multicolumn{2}{l|}{\fourj} \\
\cline{2-5}
(year)               &  data & background & data & background \\
\hline\hline
189~GeV (1998)       & 69 & $70.2\pm1.6$ & 309 & $338.9\pm3.5$\\
$192-202$~GeV (1999) & 103 & $96.1\pm1.1$ & 413 & $396.5\pm2.3$ \\
$203-209$~GeV (2000) & 159 & $150.6\pm2.7$ & 378 & $382.4\pm4.2$ \\  
\hline
\end{tabular}
\end{center}
\caption{\sl 
Observed data and expected \SM\ background events for each year
for the \twojt\ and \fourj\  final states. 
The uncertainty on the background prediction due to the limited number of
simulated events is given.
}
\label{tab:events}

\end{table}

\begin{figure}[p!] 
\centering
\epsfig{file=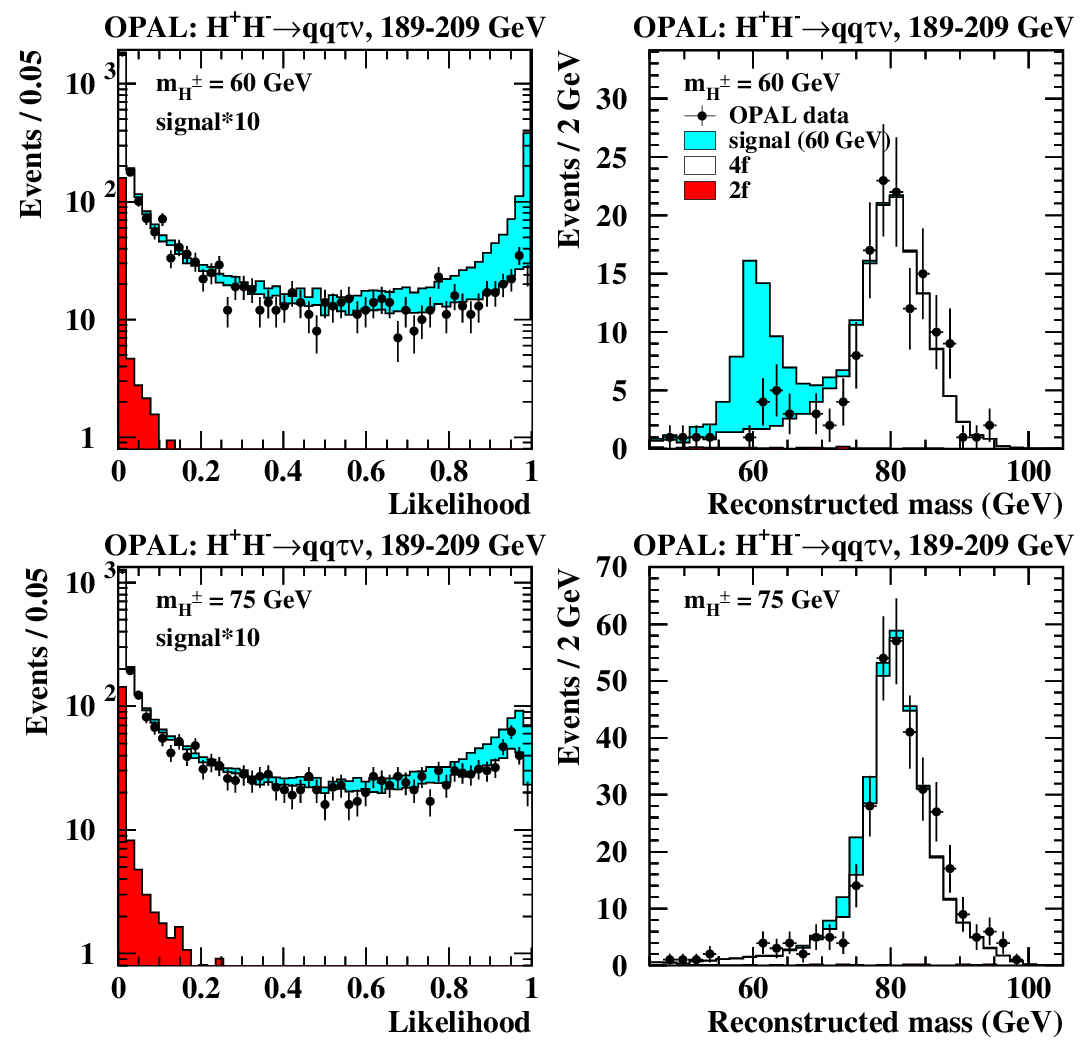,width=0.8\textwidth,height=0.545\textwidth}
\epsfig{file=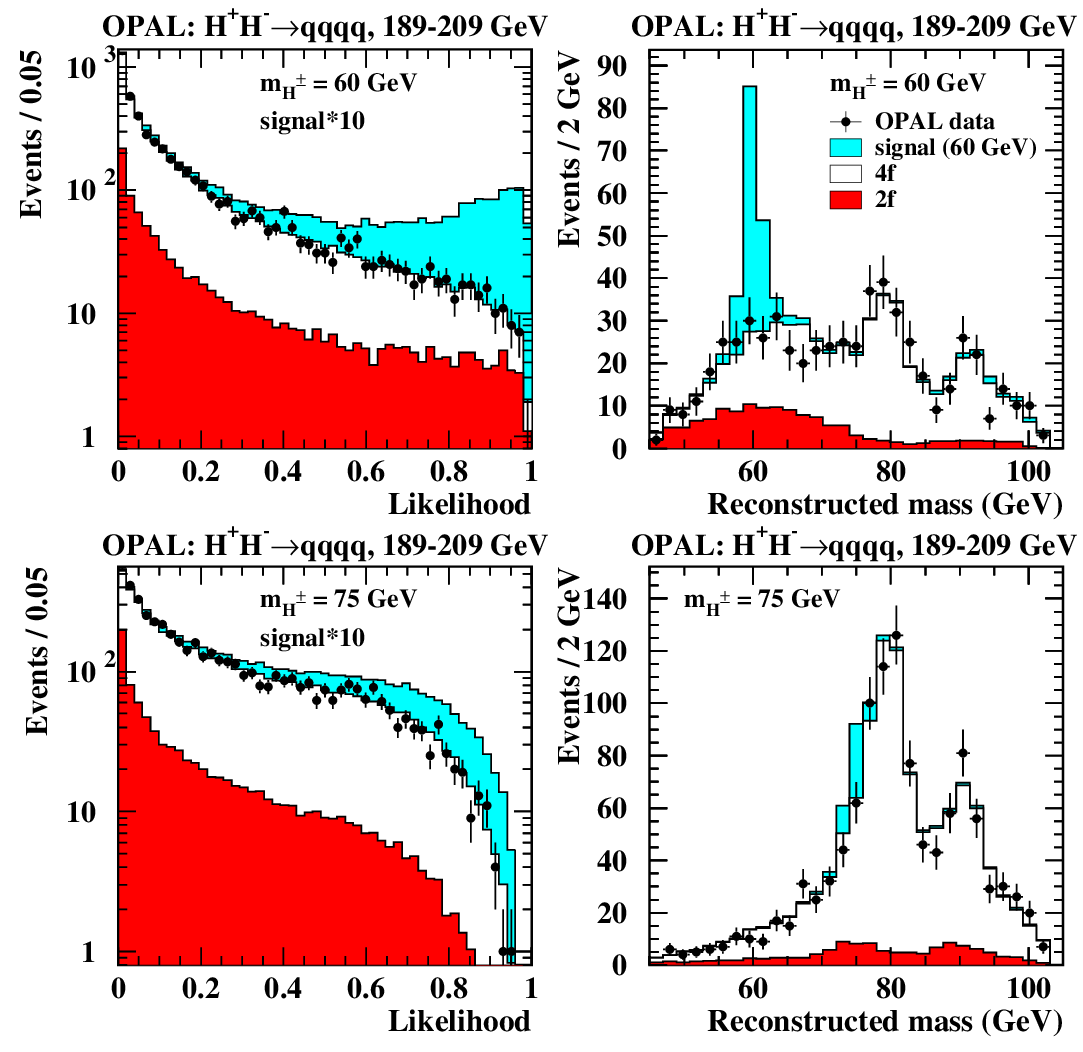,width=0.8\textwidth, height=0.545\textwidth}
\caption{\sl 
Likelihood output and reconstructed di-jet mass distributions 
for the (a-d) \twojt\ and (e-h)
\fourj\  channels. The distributions are summed up for all center-of-mass
energies and correspond 
to 60~GeV and 75~GeV simulated charged Higgs-boson masses. 
All Monte Carlo distributions are normalized to the integrated luminosity of
the data. When plotting the likelihood output, the signal expectation 
is scaled up by a factor of 10 for better visibility. A hadronic branching
ratio of 0.5 is assumed for the \twojt\ signal, and 1.0 for the \fourj\ signal. 
The reconstructed mass distributions are shown after the likelihood selection.}

\vspace*{-20.3cm}

\hspace*{4.9cm} (a) \hspace*{5.3cm} (b) \vspace*{3.75cm}

\hspace*{4.9cm} (c) \hspace*{5.3cm} (d) \vspace*{3.8cm}

\hspace*{4.9cm} (e) \hspace*{5.3cm} (f) \vspace*{3.75cm}

\hspace*{4.9cm} (g) \hspace*{5.3cm} (h) \vspace*{6cm}

\label{fig:lh}
\end{figure}

\begin{figure}[htbp!] 
\centering
\epsfig{file=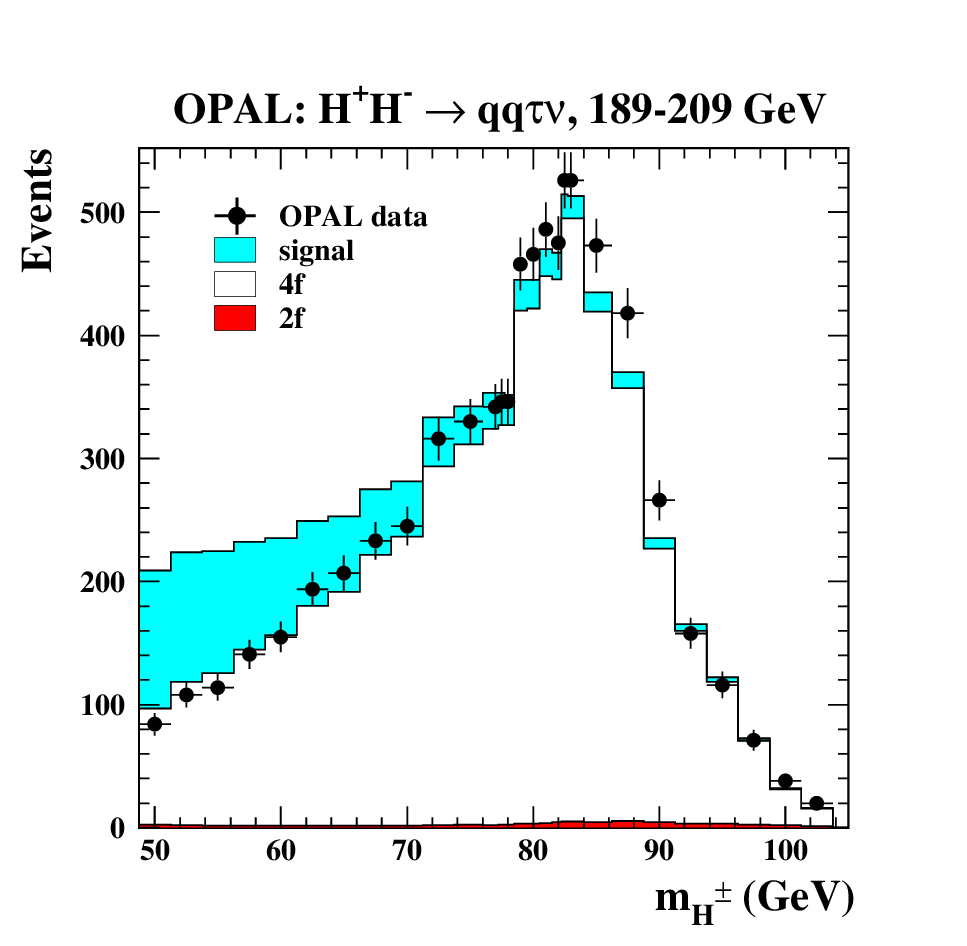,width=60mm}
\epsfig{file=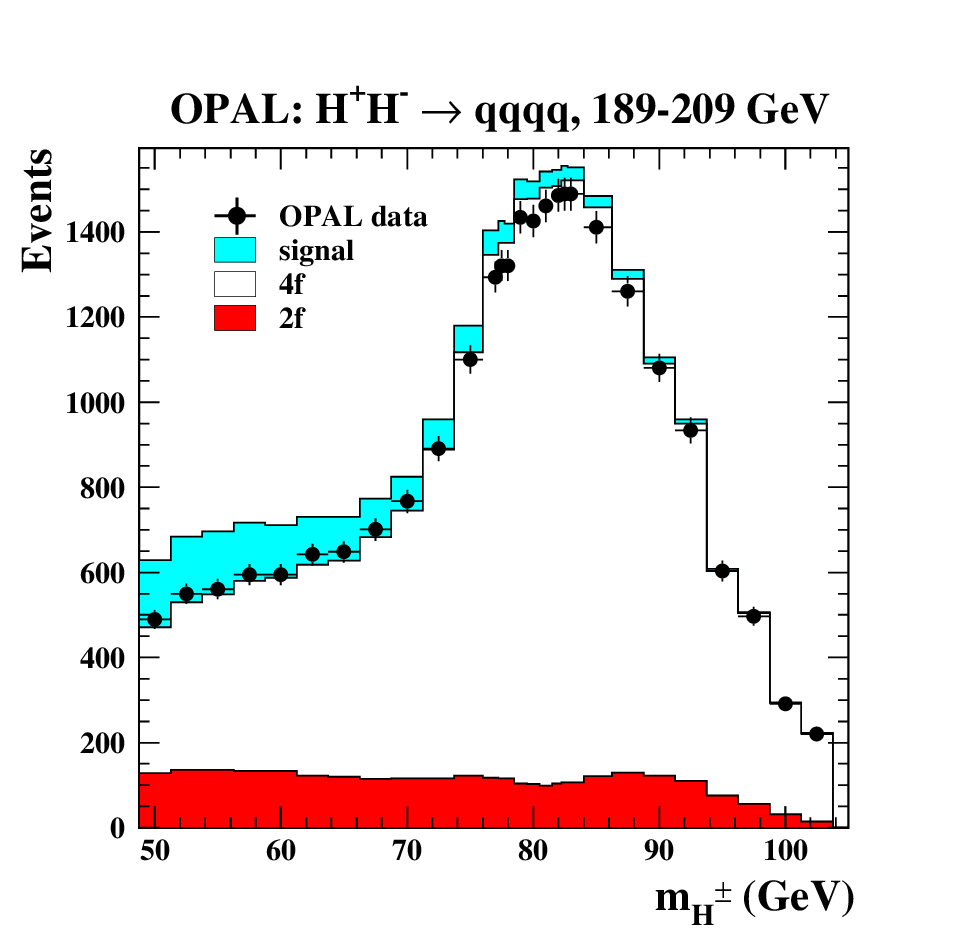,width=60mm}
\caption{\sl 
The number of observed data, expected background and signal events for the
(a) \twojt\ and (b) \fourj\ channels. The numbers are summed up for all 
center-of-mass energies 
and shown  as a function of the reference charged Higgs-boson mass.  A
hadronic branching ratio of 0.5 is assumed for the \twojt\ signal, and 1.0 for
the \fourj\
signal. Each bin corresponds to a different  likelihood selection optimized for
the mass at which the dot is centered. Since the same background simulations are
used to form the reference histograms and the same data enter the selection,
the neighboring points are strongly correlated.}

\vspace*{-8.65cm}

\hspace*{3.6cm} (a) \hspace*{5.35cm} (b) \vspace*{8.2cm}

\label{fig:mdep}
\end{figure}

The likelihood output and reconstructed di-jet mass distributions
for  simulated Higgs-boson masses of 60~GeV and 75~GeV are presented in 
Figures~\ref{fig:lh}(a-d). 
The reconstructed Higgs-boson mass resolution is $2.0-2.5$~GeV~\cite{hpaper183}. 
Figure~\ref{fig:mdep}(a) gives the mass dependence of the 
expected number of background and signal events and compares them to the
observed number of events at each test mass. 

The systematic uncertainties are estimated for several choices of the charged
Higgs-boson mass from 50~GeV to 90~GeV  at center-of-mass energies of 
\sqrts=189~GeV, 200~GeV and 206~GeV to cover the full LEP2 range. The following sources of
uncertainties are considered: limited number of generated Monte Carlo events,
statistical and systematic uncertainty on the luminosity measurement, modeling of
kinematic variables in the preselection and in the likelihood selection, tau lepton
identification, dependence of the signal detection efficiency on final-state
quark flavor, signal selection efficiency interpolation between generated Monte
Carlo points, background hadronization model, and
four-fermion background model. The contributions from the different sources are
summarized in Table~\ref{tab:syst}.

In the limit calculation, the efficiency and background estimates of the
\twojt\ channel  are reduced by 0.8$-$1.7\% (depending on the
center-of-mass energy)  in order to account for accidental vetoes  due to
accelerator-related backgrounds in the forward detectors.

\begin{table}[htbp!]
\begin{center}
\begin{tabular}{|l||r|r||r|r|}\hline
Source & \multicolumn{2}{c||}{\twojt} & \multicolumn{2}{c|}{\fourj} \\ 
\cline{2-5}
                     & signal    & background & signal    & background \\ 
\hline\hline
MC statistics        & 3.1$-$4.6 & 1.4$-$4.3  & 1.6$-$2.4 & 0.9$-$1.9 \\
luminosity           & 0.3       & 0.3        & 0.3       & 0.3 \\
preselection         & 1.5$-$4.7 & 1.8$-$7.6  & 0.3$-$1.1 & 0.5$-$2.2 \\
likelihood selection & 0.9$-$6.5 & 5.8$-$22.7 & 0.7$-$2.4 & 2.1$-$7.5 \\
tau identification   & 3.0       & 3.0        & N.A.      & N.A. \\
quark flavor        & 2.7$-$3.8 & N.A.       & 1.2$-$6.4 & N.A. \\
interpolation        & 0.2$-$0.4 & N.A.       & 0.7$-$3.7 & N.A. \\
hadronization model  & N.        & 1.0$-$2.7  & N.        & 0.7$-$4.1 \\
4f background model  & N.A.      & 0.3$-$3.3  & N.A.      & 1.7$-$3.7 \\
\hline
\end{tabular}
\end{center}
\caption{\sl
Relative systematic uncertainties on the expected background and signal rates
for the \twojt\ and \fourj\ final states. 
The numbers are given in
\% and depend on the center-of-mass energy and the reference charged
Higgs-boson mass. N.A. stands for not applicable, N. for negligible.}
\label{tab:syst}
\end{table}

%%%%%%%%% 4JET

\subsection{The four-jet final state}\label{sect:4jet}

The event selection  follows our published analysis at 
\sqrts=183~GeV~\cite{hpaper183}: 
first, well-separated four-jet events with large visible energy
are preselected; 
then a set of variables is combined using a likelihood technique.
To improve the discriminating power of the likelihood selection,
a new reference variable is introduced: 
the logarithm of the matrix element probability
for \WW\ production averaged 
over all possible jet-parton assignments
computed by EXCALIBUR~\cite{excalibur}. Moreover, 
we introduce mass-dependent likelihood functions as explained above.
As the optimal cut value on the likelihood output is not that sensitive
to the charged Higgs-boson mass in this search channel, 
we use the condition
\LH\ $>$ 0.45 at all center-of-mass energies and for all test masses.

There is a good agreement between the observed data and the \SM\ Monte Carlo
expectations at all stages of the selection. 
The number of selected events per year is given in Table~\ref{tab:events}
for a test mass of \mHpm=75~GeV.
In total, 
% After all cuts, 
1100 events are
selected in the data, while  $1117.8 \pm 5.9$~(stat.)~$\pm 74.4$~(syst.)  events
are expected from \SM\ processes.
% for a test mass of \mHpm=75~GeV.  
The four-fermion processes
account for about 90\% of the expected background and result in a large peak
centered at the \Wpm\ mass and a smaller one at the Z boson mass.   The signal
detection efficiencies are between 41\% and 59\% for any test mass  and
center-of-mass energy.  

Typical likelihood output and reconstructed di-jet mass
distributions of the selected events together with the \SM\ background
expectation and signal shapes for  simulated charged Higgs-boson masses of 60~GeV 
and 75~GeV are plotted in Figures~\ref{fig:lh}(e-h).  
The Higgs-boson mass can be reconstructed with a resolution of 
$1-1.5$~GeV~\cite{hpaper183}.   
Figure~\ref{fig:mdep}(b)
shows the mass dependence of the expected  number of background and
signal events and compares them to the  observed number of events at each test
mass. Systematic uncertainties are estimated in the same manner as for the
\twojt\ search and are given in Table~\ref{tab:syst}.

%%%%%%%%% WAWA

\section{\boldmath Search for \AWp\AWm\ events}\label{sect:wawa}

In a large part of the \THDMI\ parameter space, the branching  ratio
of \Hpm\ $\rightarrow$ \AWpm\ dominates. The possible decay modes of the A
boson and the \Wpms\  lead to many possible \HH\ra\AWp\AWm\ event topologies. 
Above \mA$\approx$12~GeV, 
the A boson decays predominantly into a \bb\ pair,  and thus its detection is
based on b-flavor identification. Two possibilities, covering 90\% of the
decays of two \Wpms, are  considered:   quark pairs from both \Wpms\ bosons or
a quark pair from one and a leptonic final state from the other. The event
topologies are therefore ``eight jets" or ``six jets and a 
lepton with missing energy", with four jets containing b-flavor in both cases.

The background comes from several Standard 
Model processes. ZZ and \WW\ production can result in multi-jet 
events. While ZZ events can contain true b-flavored jets, 
\WW\ events are selected as candidates when c-flavored jets fake b-jets. 
Radiative QCD corrections to \ee\ra\qq\ also
give a significant contribution to the expected background.

Due to the complexity of the
eight-parton final state, it is more efficient to use general event properties
and variables  designed specifically to discriminate against the main
background than a full reconstruction of the event. As a consequence, 
no attempt is made to reconstruct the charged Higgs-boson mass. 

The analysis proceeds in two steps. First a  preselection is applied  to select
b-tagged multi-jet events compatible with the signal
hypothesis. Then a likelihood selection (with three event classes: signal,
four-fermion background and two-fermion background) is applied.

The preselection of multi-jet events uses the same variables as the search for
the hadronic final state in~\cite{hpaper183} with optimized cut
positions. However, it introduces a very powerful new criterion, especially
against the \WW\ background, on a combined
b-tagging variable (\bevt)
requiring the consistency of the event with the presence of b-quark jets.

The neural network method used for b-tagging in the
OPAL \SM\ Higgs-boson search~\cite{SMhiggs} is used
to calculate on a jet-by-jet basis the discriminating 
variables $f^i_\mathrm{c/b}$ and $f^i_\mathrm{uds/b}$. 
These are constructed for
each jet $i$ as  the ratios of probabilities for the jet to be c- or uds-like
versus the probability to be b-like. The inputs to the neural network include
information about the presence of secondary vertices in a jet,
the jet shape, and the presence of leptons with large transverse
momentum.
The Monte Carlo description of the neural network output was checked with 
LEP1 data with a jet energy of about 46 GeV. 
The main background in this search at LEP2 comes from four-fermion processes,
in which the mean jet energy is about 50 GeV, very close to the LEP1 jet energy;
therefore, an adequate modeling 
%of the background is expected with an event reconstruction assuming four jets.
of the background is expected with the events reconstructed as four jets.

The \AWp\AWm\ signal topology depends on the Higgs-boson masses. 
At \mA$\approx$12~GeV or \mA$\approx$\mHpm, the available energy in the
A or \Wpms\ system is too low to form two clean, collimated jets.
At high \mHpm, the boost of the A and \Wpms\ bosons is small in the laboratory 
frame and the original eight partons cannot be identified.
At low \mHpm, the A and \Wpms\ bosons might have a boost, but it is still not
possible to resolve correctly the two partons from their decay.
From these considerations, one can conclude that it is not useful to 
require eight (or even six) jets in the event, as these jets will 
not correspond to the original partons. 
Consequently, to get the best possible modeling of the background, 
four jets are reconstructed with the Durham jet-finding
algorithm~\cite{durham} before the b-tagger is run.
 
The flavor-discriminating variables are combined for the four
reconstructed jets by 
\begin{equation} 
\bevt =
\frac{1}{1+\alpha \cdot \prod_i f^i_\mathrm{c/b}  +
\beta  \cdot \prod_i f^i_\mathrm{uds/b}}
\label{equ:bevt} \end{equation} 
The index $i$ runs over the reconstructed jets ($i=1,...4$) and 
the parameters $\alpha$ and $\beta$ are numerical coefficients whose optimal
values depend on the flavor composition of the signal and background final
states. However, since the
expected sensitivity of the search is only slightly dependent on the values of 
$\alpha$ and $\beta$, they are fixed at $\alpha=0.1$ and $\beta=0.7$. Events are
retained if $\bevt>0.4$.

The preselections of the two event topologies (\eightj\ and \sixjl) are very 
similar.  However, in the \sixjl\
channel, no kinematic fit is made to the \WW\ra\qq\qq\ hypothesis and,
%therefore, no cuts are made on the fit probabilities. No lepton
therefore, no cuts are made on the fit probabilities. No charged lepton
identification is applied;  instead the search is based on indirect detection
of the associated neutrino by measuring the missing energy. 

After the preselection the observed data show an excess over the predicted Monte Carlo
background. This can partly be explained by the apparent difference between the
gluon splitting rate into c\=c and b\=b pairs in the data and in the background
Monte Carlo simulation. The measured rates at \sqrts$=$91~GeV are
$g_\mathrm{c\bar{c}} = 3.2 \pm 0.21 \pm 0.38\%$~\cite{gtocc} and 
$g_\mathrm{b\bar{b}} = 0.307 \pm 0.053 \pm 0.097\%$~\cite{gtobb} 
from the LEP1 OPAL data. 
The gluon splitting rates in our Monte Carlo simulation are extracted from
%\ee\ra\ZZ\ra\lplm\qq\ events and are
\ee\ra\ZZ\ra\lplm\qq\ events,
where the Z\ra\qq\ decays have similar kinematic properties to the ones in the
LEP1 measurement.
Note that \ee\ra\ZZ\ra\qq\qq\ events can not be used as the 
%%since the hadronic decay of both Z's leads to 
two \qq\ pairs interact strongly
with each other. 
%%They are found to be 
The rates are found to be
$g_\mathrm{c\bar{c}}^\mathrm{MC} = 1.33 \pm 0.06\%$ and 
$g_\mathrm{b\bar{b}}^\mathrm{MC} = 0.116 \pm 0.0167\%$, 
averaged over all center-of-mass energies.
This mismodeling can be compensated by reweighting the SM Monte Carlo 
events with gluon splitting to heavy quarks 
% by universal reweighting factors~\cite{gluon-split-theory} 
and at the same time 
deweighting the non-split events
to keep the total numbers of \WW, \ZZ\ and two-fermion 
background events fixed at generator level. 
The reweighting factor is 2.41 for g\ra c\=c and 2.65 for g\ra b\=b.
The same reweighting factors are used for \WW, \ZZ\ and two-fermion 
events with gluon splitting at all LEP2 energies,
noting that all background samples were hadronized with the same settings and  
assuming that the \sqrts\ dependence of the gluon splitting of a fragmenting
two-fermion system is correctly modeled by the Monte Carlo generator.
It is known that the generator reproduces the energy dependence predicted by
QCD in the order $\alpha_\mathrm{s}$ with resummed leading-log and
next-to-leading log terms~\cite{gluon-split-theory}.
This correction results in a background enhancement factor of 
1.08 to 1.1 after the preselection, 
depending on the search channel and the center-of-mass energy, but
it does not affect the shape of the background distributions.

\begin{figure}[p!]
\centering
\vspace*{-0.52cm}

\epsfig{file=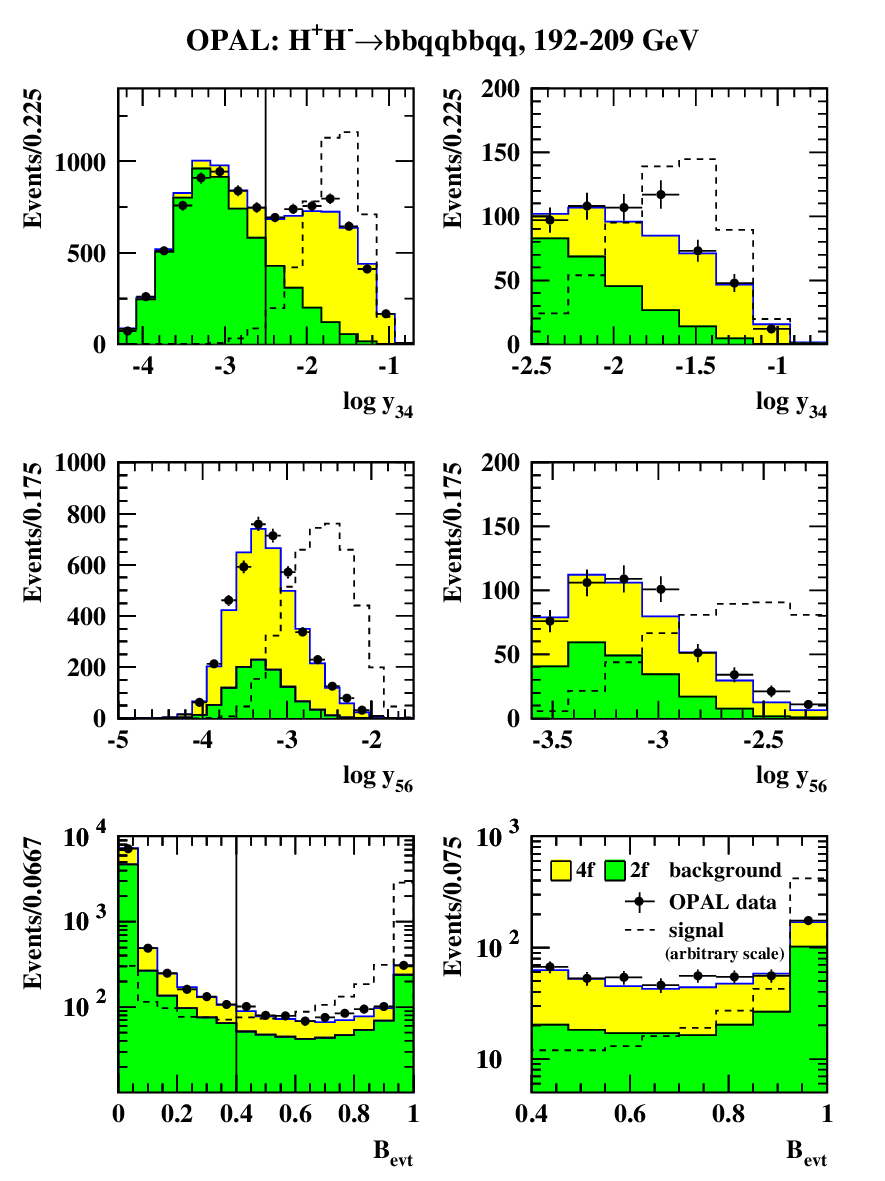, width=0.94\textwidth}
\caption{\sl 
Most important selection variables: (a-b) $\log_{10} y_{34}$, (c-d)
$\log_{10} y_{56}$ and (e-f) \bevt\
in the \eightj\ channel at
$\sqrts=192-209$~GeV. The distributions are shown 
(left) in a background-enriched data sample (see text for explanation)
and (right) after the full preselection. 
To form the signal histograms, the Monte Carlo distributions
are averaged for all simulated \mHmAc\ mass combinations in the mass range of
interest.
The Monte Carlo reweighting to the measured gluon splitting rates is included. 
The expectations from \SM\ processes are
normalized to the data luminosity. 
The preselection cuts on $y_{34}$ and \bevt\
are indicated by vertical lines. 
}

\vspace*{-22.1cm}

\hspace*{4.9cm} (a) \hspace*{6.1cm} (b) \vspace*{5.7cm}

\hspace*{4.9cm} (c) \hspace*{6.1cm} (d) \vspace*{5.6cm}

\hspace*{4.9cm} (e) \hspace*{6.1cm} (f) \vspace*{8cm}

\label{fig:8jvars}
\end{figure}

The numbers of preselected events after the reweighting 
%are given in Table~\ref{tab:wawa_sel}.
are given in columns 2 and 3 of Table~\ref{tab:wawa_sel}.
At this stage of the analysis the \eightj\ and \sixjl\ data samples are 
highly overlapping. The observed rates still show an excess over the 
background predictions, adding up to about 1.6 standard deviations 
%%for either preselection.
in both samples.
Although this difference is statistically not significant, it
can be shown that the Monte Carlo prediction has minor imperfections.
For the \eightj\ case, the distributions of three variables used in
the analysis, namely $y_{34}$, $y_{56}$ and $\mathcal{B}_{\rm evt}$,
are plotted in the right part of Figure~\ref{fig:8jvars}.
As can be seen, the variable $y_{56}$ is most powerful to reject the
background. Both the $y_{34}$ and the $y_{56}$ distributions are slightly
shifted towards the position of a hypothetical Higgs signal and
the  $\mathcal{B}_{\rm evt}$ distribution shows an excess over the 
predicted background at intermediate $\mathcal{B}_{\rm evt}$ values, but 
the excess events are not distributed according to the expectation 
for a Higgs signal.
The shifts are visible with better statistical significance in the left 
part of Figure~\ref{fig:8jvars}. It shows the same variables for a background 
enriched data sample, where the preselection cuts on $y_{34}$ and 
$\mathcal{B}_{\rm evt}$ are dropped, except for the study of the 
$y_{56}$ variable where we keep the cut on $y_{34}$ to select multi-jet 
events. The resulting samples are completely dominated by background, 
the contribution of a Higgs signal being at most 0.5\%. Since heavy 
quark production in the Monte Carlo generator is already corrected, 
the origin of the discrepancies is likely a slight mismodeling of
the topology of multi-jet events, especially if they contain
heavy quarks. 
No further correction is applied to the estimated background.
Excess events passing the final selection, 
even if they do not look signal-like, are thus counted 
with a certain weight as signal events in the statistical analysis, to be 
discussed later.

\begin{table}[t!]
\begin{center}
\begin{tabular}{|l||l|l||l|l|l|}\hline
LEP energy        & preselection & preselection & exclusive & exclusive & overlap \\
(year)            & \eightj& \sixjl    & \eightj   & \sixjl    &  \\
\hline\hline
189~GeV     \hfill data     & 238    &  358      &  3        & 24        & 5    \\
(1998) \hspace*{1.2cm} background & 231.2$\pm$2.9  
                            &  342.2$\pm$3.6    
			                &  2.1$\pm$0.3      
					            & 24.4$\pm$1.0      
						                & 6.3$\pm$0.5 \\
\hline
$192-202$~GeV \hfill data   & 297    &  310      & 16        & 16        & 17   \\
(1999) \hspace*{1.2cm} background & 270.4$\pm$2.9  
                            &  285.0$\pm$3.0    
			                & 13.3$\pm$0.7      
					            & 10.4$\pm$0.6       
						                & 13.4$\pm$0.7 \\
\hline
$200-209$~GeV \hfill data   & 265    &  281      &  9        & 8         & 15 \\
(2000) \hspace*{1.2cm} background & 252.5$\pm$3.7  
                            &  270.5$\pm$5.0    
			                & 13.0$\pm$0.9      
					            & 9.3$\pm$0.8       
						                & 12.9$\pm$0.9 \\  
\hline
\end{tabular}
\end{center}
\caption{\sl 
Observed data and expected \SM\ background events for each year
in the \AWp\AWm\ searches. The \eightj\
and \sixjl\ event samples after the preselection step (2nd and 3rd columns)
are highly overlapping.
After the likelihood selection, the overlapping events are removed from the
\eightj\
and \sixjl\ samples and form a separate search channel (last three columns).
The uncertainty on the background prediction due to the limited number of
simulated events is given.
The Monte Carlo reweighting to the measured gluon splitting rates is included.
}
\label{tab:wawa_sel}
\end{table}

As a final selection, likelihood functions are built to identify signal events.
The reference distributions depend on the LEP energy, but they 
are constructed to be independent of the considered
\mHmAc\ combination. To this end, we form the signal reference distributions
by averaging all simulated \HH\ samples 
in the \mHmAc\ mass range of interest. 

Since the
selections at $\sqrts=192-209$~GeV are aimed at charged Higgs-boson masses
around the expected sensitivity reach of about 80$-$90~GeV, 
all masses up to
the kinematic limit are included. On the other hand,  
at \sqrts=189~GeV only charged Higgs-boson masses up to 50~GeV are 
included since the selections at this energy 
are optimized to reach down to as low as a charged
Higgs-boson mass of 40~GeV where the LEP1 exclusion limit lies.
The input variables for the \eightj\ final state are:
the Durham jet-resolution parameters\footnote{
Throughout this paper $y_{ij}$ denotes the parameter of the Durham jet
finder at which the event classification changes from $i$-jet to
$j$-jet, where $j=i+1$.} $\log_{10} y_{34}$ and $\log_{10} y_{56}$,  
the oblateness~\cite{oblateness} event shape variable,
the opening angle of the widest jet defined by the size of the 
cone containing 68\% of the total jet energy,
%the charge-signed cosine of the production angle in the \WW\ra\qq\qq\ 
%hypothesis,
the cosine of the W production angle multiplied with the W charge 
(calculated from the jet charges~\cite{jetcharge}) for the 
e$^+$e$^- \rightarrow$ W$^+$W$^- \rightarrow q \overline{q} q 
\overline{q}$ \ interpretation, 
% as obtained from a kinematic 
% 5C fit~\cite{5Cfit} and from the jet charges~\cite{jetcharge},
and the b-tagging variable \bevt.
At \sqrts=189~GeV,
$\log_{10} y_{23}$, $\log_{10} y_{45}$, $\log_{10} y_{67}$, 
and the maximum jet energy  are also used. Moreover, 
the sphericity~\cite{sphericity} event shape variable has more discriminating
power and thus replaces oblateness.
Although the $y_{ij}$ variables are somewhat correlated, they contain additional
information: their differences reflect the kinematics of the initial partons.

The input variables for the \sixjl\ selection are:
$\log_{10} y_{34}$, $\log_{10} y_{56}$, the oblateness, 
the missing energy of the event, 
and \btagvar.
At \sqrts=189~GeV, $\log_{10} y_{23}$, the maximum jet energy
and the sphericity are also included.

\begin{figure}[htbp!]
\centering
\epsfig{file=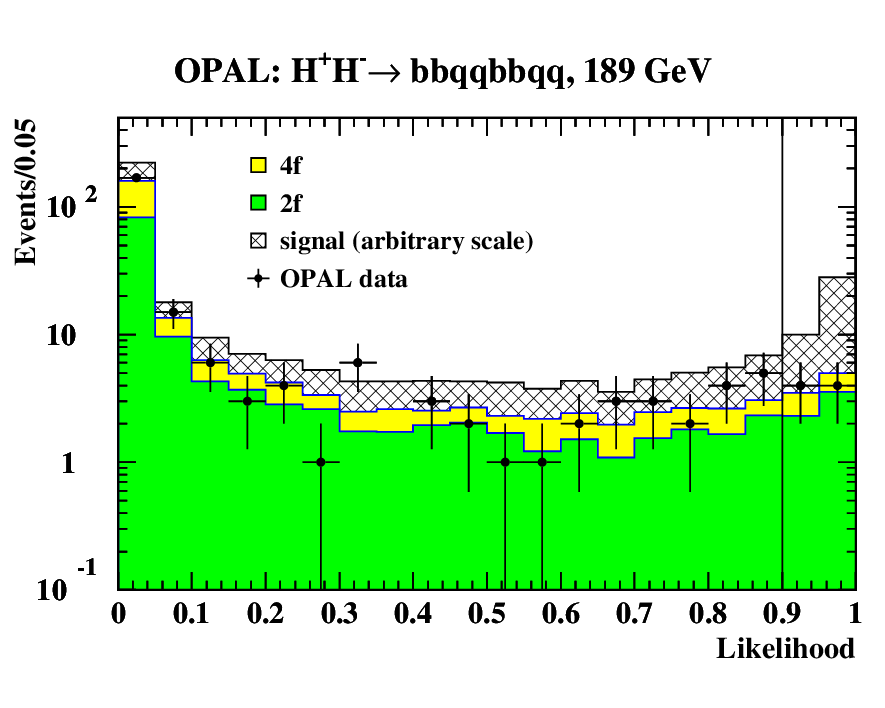, width=3. in}
\epsfig{file=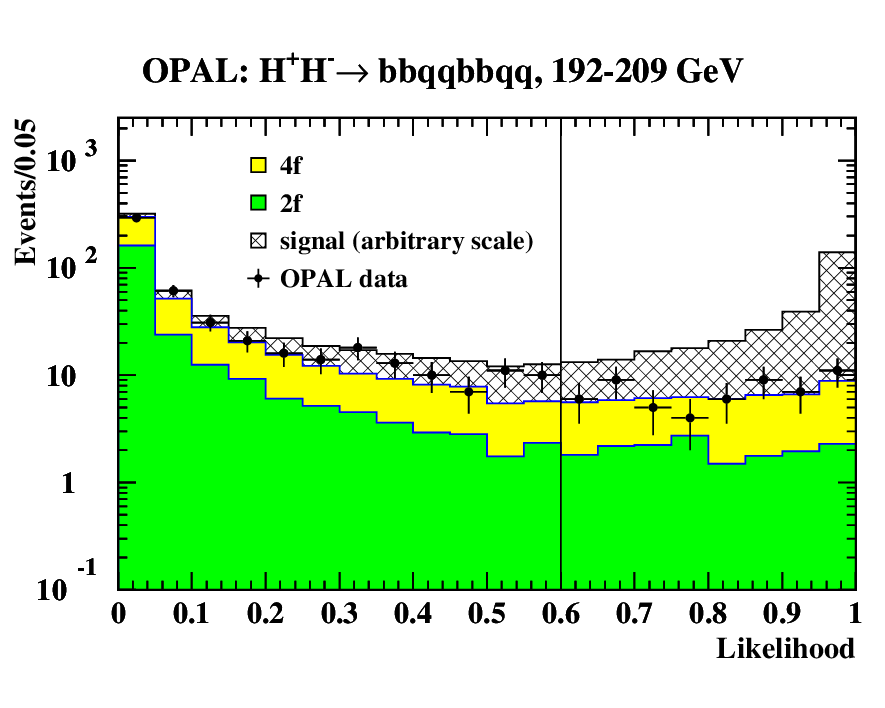, width=3. in}

\epsfig{file=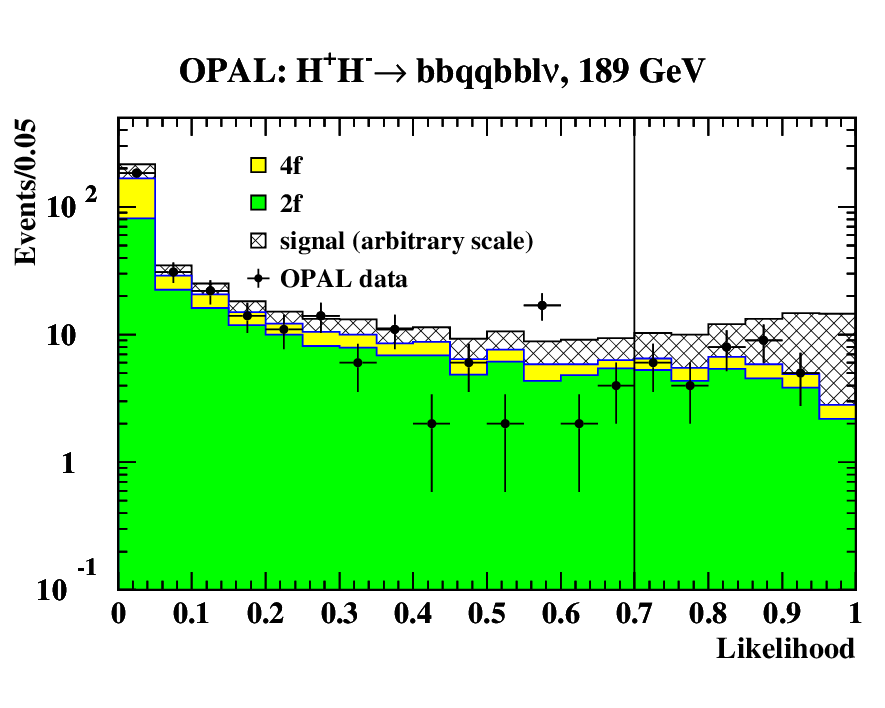, width=3. in}
\epsfig{file=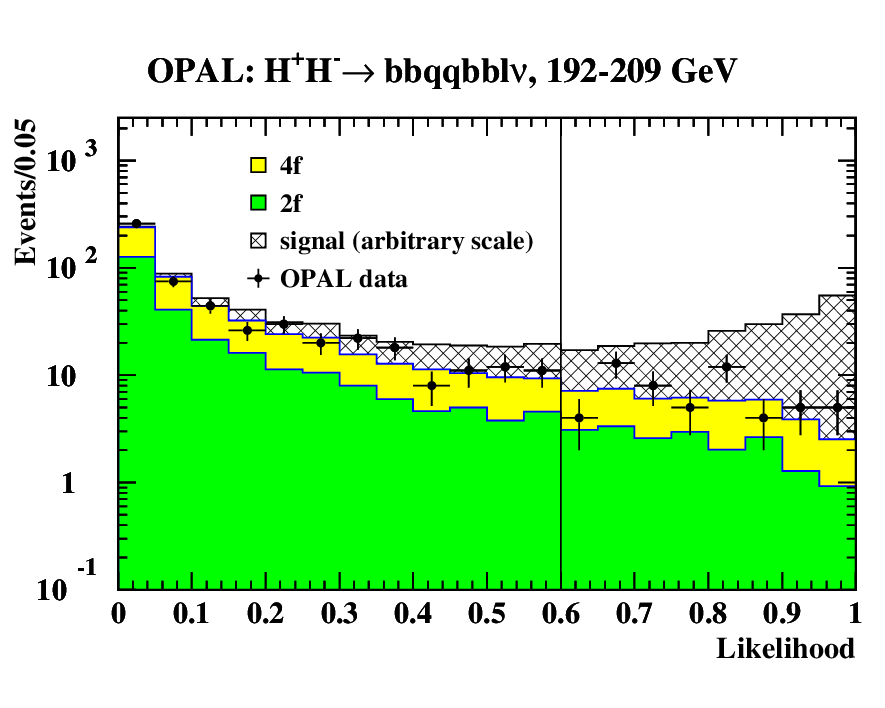, width=3. in}
\caption{\sl 
Likelihood output distributions for the (a-b) \eightj\ and 
(c-d) \sixjl\ channels  at
\sqrts=189~GeV and  $192-209$~GeV. 
To form the signal histograms, the Monte Carlo distributions
are averaged for all simulated \mHmAc\ mass combinations in the mass range of
interest. 
The Monte Carlo reweighting to the measured gluon splitting rates is included.
The expectations from  \SM\ processes are
normalized to the data luminosity. The lower cuts on the likelihood output are
indicated by vertical lines. 
}

\vspace*{-14.4cm}

\hspace*{6.3cm} (a) \hspace*{6.9cm} (b) \vspace*{5.7cm}

\hspace*{6.3cm} (c) \hspace*{6.9cm} (d) \vspace*{7.5cm}

\label{fig:wawalik}
\end{figure}

Events are selected if they pass a lower cut on the likelihood output.
%This final cut does not remove  (especially in the year 1999
%data) all the excess events observed after the preselection.
%The distributions of 
% the critical selection variables, 
%$y_{34}$, 
%and the most powerful variables to reduce the background,
%$y_{56}$ and \btagvar, are plotted in 
%Figure~\ref{fig:8jvars}
%both in background-enriched data samples and after the preselection.
%To prepare the background-enriched data samples, the preselection cuts on $y_{34}$ and 
%\btagvar\ are dropped, except for the study of the $y_{56}$
%variable where we keep the cut on $y_{34}$ in order to select multi-jet events.
%The resulting samples are completely dominated by background processes. 
%We find systematic differences between the data samples and 
%the equivalent background Monte Carlo simulations at both stages. 
%The observed excess  
%cannot be attributed to a Higgs-boson signal, because 
%its distributions do not agree with the predictions for a signal, and the
%contribution of a signal in the background enriched data sample
%is expected to be at most 0.5\%.
%The interpretation of the excess in terms of a
%systematic uncertainty on the background prediction will be discussed
%later.
The likelihood distributions are shown in Figure~\ref{fig:wawalik}. 
The positions of the likelihood cuts are indicated by vertical lines.
The discrepancies observed in Figure~\ref{fig:8jvars} in background-enriched
samples,  
propagate into the likelihood distributions. 
Since the excess events in Figure~\ref{fig:8jvars} are shifted relative to 
the background
expectation, but do not agree with the Higgs distribution, they give
likelihood values between the mean background and signal values in 
Figure~\ref{fig:wawalik}.
%On average, the observed likelihood
%values are
%higher than predicted for background but lower
%than the expected for signal events. 
With large statistical errors,
the effect can be seen at intermediate likelihood values.
Some of the excess events pass the final likelihood cut.

%The non-signal nature of the excess is also demonstrated by 
%Figure~\ref{fig:wawalik}, which shows the distributions of the likelihood output.
%The positions of the likelihood cut are indicated by vertical lines.

%Because the \eightj\ and the \sixjl\ selections have several discriminating
%variables in common, a major overlap between these selections exists. To 
%assure that every event is counted only once, the two samples are 
%redistributed into three: ($i$) events exclusively classified as \eightj\ 
%candidates, ($ii$) events
To assure that every event is counted only once in the final analysis,
the overlapping \eightj\ and \sixjl\ event samples, as obtained after the 
final likelihood cut are redistributed into three event classes:
($i$) events exclusively classified as \eightj\ candidates, ($ii$) events
exclusively classified as \sixjl\ candidates and ($iii$) events accepted by both
selections. If an event falls into class ($iii$), the larger likelihood output of
the two selections  is kept for further processing.  The final results using
the above classification are quoted in Table~\ref{tab:wawa_sel}.
After all selection cuts, an excess of events appears in the
1999 data sample. The excess ($1.9 \sigma$) is not statistically significant 
and 
%%statistically 
it is consistent with the results 
%%for 
of the other years.
This modified channel definition not only removes the overlap but also 
increases the efficiency for detecting signal events by considering the
cross-channel efficiencies  (e.g. the efficiency to select \HH\ra\bb\qq\bb\qq\ 
signal by the exclusive \sixjl\ selection can be as high as 18\%, though it is
typically only a few \%). The efficiencies are determined independently
for all simulated \mHmAc\  combinations and interpolated to arbitrary \mHmAc\
by two-dimensional spline interpolation. 
The behavior of the selection
efficiencies depends strongly on the targeted charged Higgs-boson mass range and also 
varies with the mass difference  $\Delta m=\mHpm-\mA$.
In most cases the overlap channel has the highest efficiency. At \sqrts=189 GeV
and \mHpm=45~GeV, it reaches 32\% for the \bb\qq\bb\lnl\ and 
44\% for the \HH\ra\bb\qq\bb\qq\ signal.
%% close to the \mHmAc\ diagonal (!!! NOT TRUE).
At \sqrts=206~GeV and \mHpm=90 GeV, the overlap efficiency can be as high as 62\%
for the \bb\qq\bb\lnl\ and 71\% for the \HH\ra\bb\qq\bb\qq\ signal.
The exclusive \sixjl\ selection has efficiencies typically below 20$-$30\%, while
the exclusive \eightj\ selection below 10$-$15\%. 
Table~\ref{tab:effi-WA} gives the selection efficiencies at selected \mHmAc\
points. 

%%% Running the efficiency code in 2012 Jan:
%%% 8j
%%% mH=45, mA=26: 44%
%%% mH=45, mA=42: 19%
%%% 6j+l
%%% mH=45, mA=28: 32%
%%% mH=45, mA=42: 15%

\begin{table}[htbp!]
\begin{center}
\begin{tabular}{|l||l||r|r||r|r|}\hline
 & & \multicolumn{4}{c|}{\mHmAc\ (GeV, GeV)} \\
\cline{3-6}
signal & selection & (45,30) & (80,50) & (45,30) & (90,60) \\ 
\cline{3-6}
 & & \multicolumn{2}{c||}{\sqrts=189~GeV} 
   & \multicolumn{2}{c|}{\sqrts=206~GeV} \\ 
\hline\hline
\bb\qq\bb\qq & \eightj \phantom{\LARGE T} & 4.6 & 1.0 & 9.3 & 12.4 \\
             & overlap & 41.0 & 2.8 & 14.9 & 69.6 \\
	     & \sixjl & 17.0 & 3.6 & 4.1 & 3.1 \\
\cline{2-6}
	     & total & 62.6 & 7.4 & 28.3 & 85.0 \\
\hline
\bb\qq\bb\lnl & \sixjl \phantom{\LARGE T} & 28.2 & 6.0 & 11.6 & 7.0 \\
              & overlap & 31.8 & 3.6 & 14.2 & 62.2 \\
	      & \eightj & 1.8 & 0.1 & 2.2 & 6.1 \\
\cline{2-6}
	      & total & 61.8 & 9.7 & 28.0 & 75.3 \\
\hline
\bb\qq\tnt & \fourjt \phantom{\LARGE T} & 68.0 & 0.0 & 12.3 & 11.1 \\
\hline
\end{tabular}
\end{center}
\caption{\sl
Signal selection efficiencies in percent for the \Hpm\ra\AWpm\ final states 
in the different search channels at \sqrts=189 and 206~GeV
at representative \mHmAc\ points. 
}
\label{tab:effi-WA}
\end{table}

The composition of the background depends on the targeted Higgs-boson mass region.
In the low-mass selection (\sqrts=189~GeV) that is optimized for 
\mHpm=40$-$50~GeV, the Higgs bosons are boosted and therefore the final state is
two-jet-like with the largest background contribution
coming from two-fermion processes:  they account for 52\% in the exclusive \eightj,
80\% in the exclusive \sixjl\ and 76\% in the overlap channel. 
On the other hand, in the high-mass analysis 
($\sqrts=192-209$~GeV) the
%four-fermion fraction is dominant: it is
four-fermion fraction is dominant:
69\% in the \eightj, 56\% in the \sixjl\ and 70\% in the overlap
channel.

Systematic errors arise from uncertainties in the preselection and
from mismodeling of the likelihood function.
The variables $y_{34}$ and \bevt\
appear both in the preselection cuts and in the likelihood definition.
The total background rate is known to be underestimated after the
preselection step. The computation of upper limits on the production
cross section, with this background rate subtracted, results in
conservative limits, assuming the modeling of the other preselection 
variables and the signal and background likelihoods to be correct. 
Therefore, no systematic uncertainty is assigned to the percentage 
of events passing the $y_{34}$ and \bevt\ preselection cuts.
The systematic errors related to preselection variables other than 
$y_{34}$ and \bevt, evaluated from background enriched 
data samples, are taken into account.

As already mentioned, the discrepancies shown in Figure~\ref{fig:8jvars} 
have an impact on the likelihood function. Event-by-event correction 
routines for the variables  $y_{34}$ and \bevt\ were 
developed to describe the observed shapes, keeping the normalization
above the preselection cuts fixed. The systematic errors were estimated by 
computing the likelihood for all MC events with the modified values of 
$y_{34}$ and \bevt\ and counting the accepted MC events.
The systematic errors related to all other reference variables were
estimated in the same manner.

Systematic uncertainties also arise due to the gluon splitting correction. 
The experimental uncertainty on the gluon splitting rate translates into
uncertainties on the
total background rates. Moreover, there is an uncertainty due to the 
Monte Carlo statistics of the g$\ra$c\=c and b\=b events.

Finally, uncertainties due to the limited number of simulated 
signal and background events are included.
%In summary, the following sources of systematic uncertainties are considered: 
%limited number of
%simulated signal and background events, modeling of the preselection
%variables (other than $y_{ij}$ and \bevt), modeling of the
%shapes of the reference distributions in the likelihood selection and the gluon
%splitting correction.
The different contributions are summarized in
Table~\ref{tab:syst-wa}.  
Uncertainties below the 1\% level are neglected. 

\begin{table}[hbtp!]
\begin{center}
{\small
\begin{tabular}{|l||r|r||r|r||r|r|}\hline
Source & \multicolumn{2}{c||}{exclusive \eightj} 
       & \multicolumn{2}{c||}{exclusive \sixjl} & \multicolumn{2}{c|}{overlap} \\ 
\cline{2-7}
                & signal    & background & signal   & background & signal   & background \\ 
\hline\hline
MC statistics   & $\ge$15/$\ge$8.5 
                            & 13.2/6.7 & $\ge$5.7/$\ge$8.4 
			                            & 4.0/8.3 & $\ge$4.0/$\ge$2.8 
						                            & 7.9/7.1 \\
\hline
preselection    & 1.0       & 1.0       & 1.0       & 2.0       & 1.0       & 2.0/1.5 \\
\hline
\LH\ selection & & & & & & \\
\hfill $y_{ij}$ & 4.0/1.8 & 6.0/6.2 & 2.2/2.6 & 6.0/8.0 & 2.8/1.5 & 5.5/4.9 \\
\hfill b-tag    & 0.0/1.8 & 4.7/7.0 & 2.9/1.4 & 4.4/7.1 & 1.0/1.3 & 4.0/5.1 \\
\hfill other    & 1.8/0.7 & 3.9/3.2 & 1.0/1.6 & 3.7/3.5 & 1.2/0.7 & 3.4/2.4 \\
\hline
gluon splitting &  &  &  &  &  &  \\
\hfill g$\ra$c\=c, exp.        & N.A.      & 0.6/2.5  & N.A.      & 1.6       & N.A.       & 1.4/2.3 \\ 
\hfill g$\ra$c\=c, MC          & N.A.      & 0.2/0.8  & N.A.      & 0.6       & N.A.       & 0.5/0.8 \\ 
\hfill g$\ra$b\=b, exp.        & N.A.      & 1.4/4.2  & N.A.      & 3.8/4.3   & N.A.       & 5.5/5.4 \\ 
\hfill g$\ra$b\=b, MC          & N.A.      & 0.5/1.7  & N.A.      & 1.5/1.7   & N.A.       & 2.2 \\ 
\hline\hline
gluon splitting &  &  &  &  &  &  \\
correction factor & N.A. & 1.05/1.18 & N.A. & 1.13/1.15 & N.A.    & 1.15/1.22 \\
\hline
\end{tabular}
}
\end{center}
\caption{\sl
Relative systematic uncertainties in percent for the \AWp\AWm\ searches. 
Where two
values are given separated by a "/",  the first belongs to the 189~GeV
selection and the second to the $192-209$~GeV selections. 
For the signal, the uncertainties due to the limited Monte Carlo statistics
are calculated by binomial statistics for a sample size of
500 events and they also depend, via the selection efficiency, on the assumed
Higgs-boson masses.
N.A. stands for not applicable.
The multiplicative gluon splitting correction factors, used to obtain the
background-rate estimates as explained in the text, are given in the last line. 
}
\label{tab:syst-wa}
\end{table}

\section{\boldmath Search for \AWpm\tnt\ events}\label{sect:WAtau}

In some parts of the \THDMI\ parameter space,  both the fermionic
\Hpm\ra\tnt\ and the bosonic \Hpm\ra\AWpm\ decay modes contribute.    To cover
this transition region 
%parallel to the \mHmAc\ diagonal
at small \mHpm$-$\mA\ mass differences, a search for the final
state \HH\ra\AWpm\tnt\ is performed. The transition region is wide for
small \tanb\ and narrow for large \tanb; therefore, this analysis is more
relevant for lower values of \tanb.  

Only the hadronic decays of \Wpms\ and the decay A\ra\bb\ are considered. Thus 
the events contain a tau lepton, four jets (two of which are b-flavored)
and missing energy. Separating the signal from the \WW\ background becomes
difficult close to \mHpm=\mWpm. 

The preselection is designed to identify hadronic events containing a tau lepton
plus significant missing energy and transverse momentum from the undetected
neutrino. In most cases it is not practical to reconstruct the four jets
originating from the \AWpm\ system. Instead, to suppress the main background  from
semi-leptonic \WW\ events, we remove the decay products of the tau candidate and 
force the remaining hadronic system into two jets by the Durham algorithm. The
requirements are then based on the preselection of Section~\ref{sect:2jet-tau}
with additional preselection cuts on the effective center-of-mass energy, 
$\log_{10}y_{12}$ and $\log_{10}y_{23}$ of the hadronic system, and 
the charge-signed \Wpm\ production angle.

The likelihood selection uses seven variables:  the momentum of the tau
candidate,  the cosine of the angle between the tau momentum and the nearest
jet, $\log_{10}y_{12}$ of the hadronic system, the cosine of the angle between
the two hadronic jets, the charge-signed cosine of the \Wpm\ production angle,
the invariant mass of the hadronic system, and the b-tagging variable \bevt. 
Here, \bevt\ is defined using the two jets of the hadronic system
using Eq. (1) of Section~\ref{sect:wawa}, with  $i=1,2$ and $\alpha=\beta=1$. 
To form the signal reference distributions, all
simulated \HH\ samples in the \mHmAc\ mass range of interest are summed up.
Since the search at $\sqrts=192-209$~GeV targets intermediate
charged Higgs-boson masses (60$-$80~GeV), 
all masses up to the kinematic limit are included. 
At \sqrts=189~GeV, 
only charged Higgs-boson masses up to 50~GeV are included since
the selection is optimized for low charged
Higgs-boson masses (40$-$50~GeV). 

\begin{figure}[htbp!]
\centering
\epsfig{file=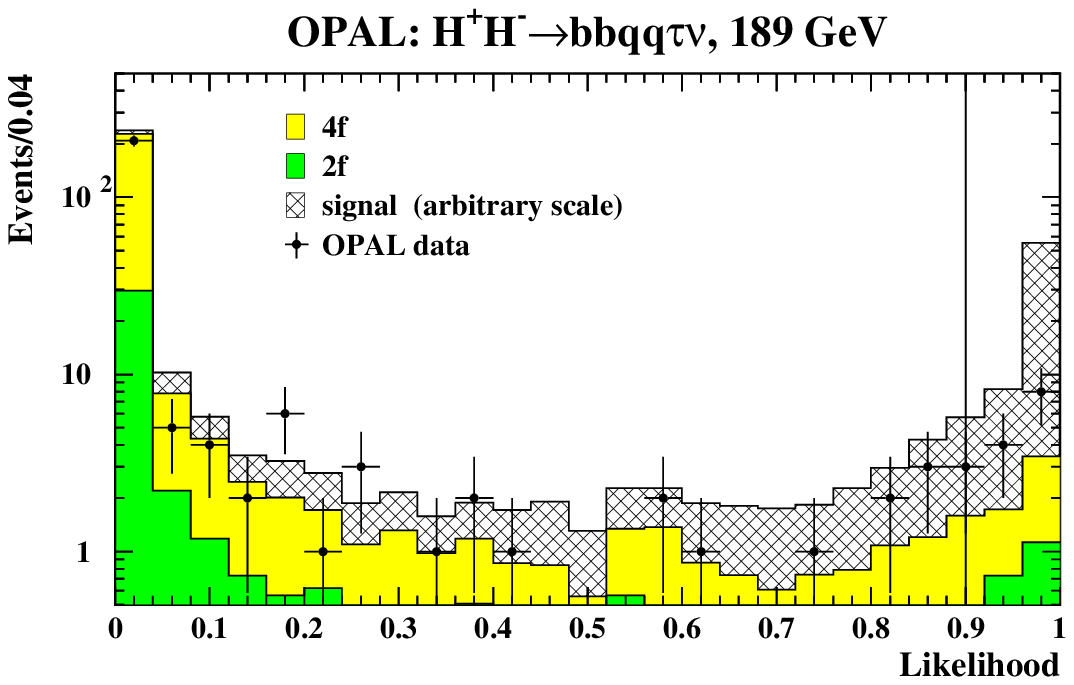, width=3.05 in}
\epsfig{file=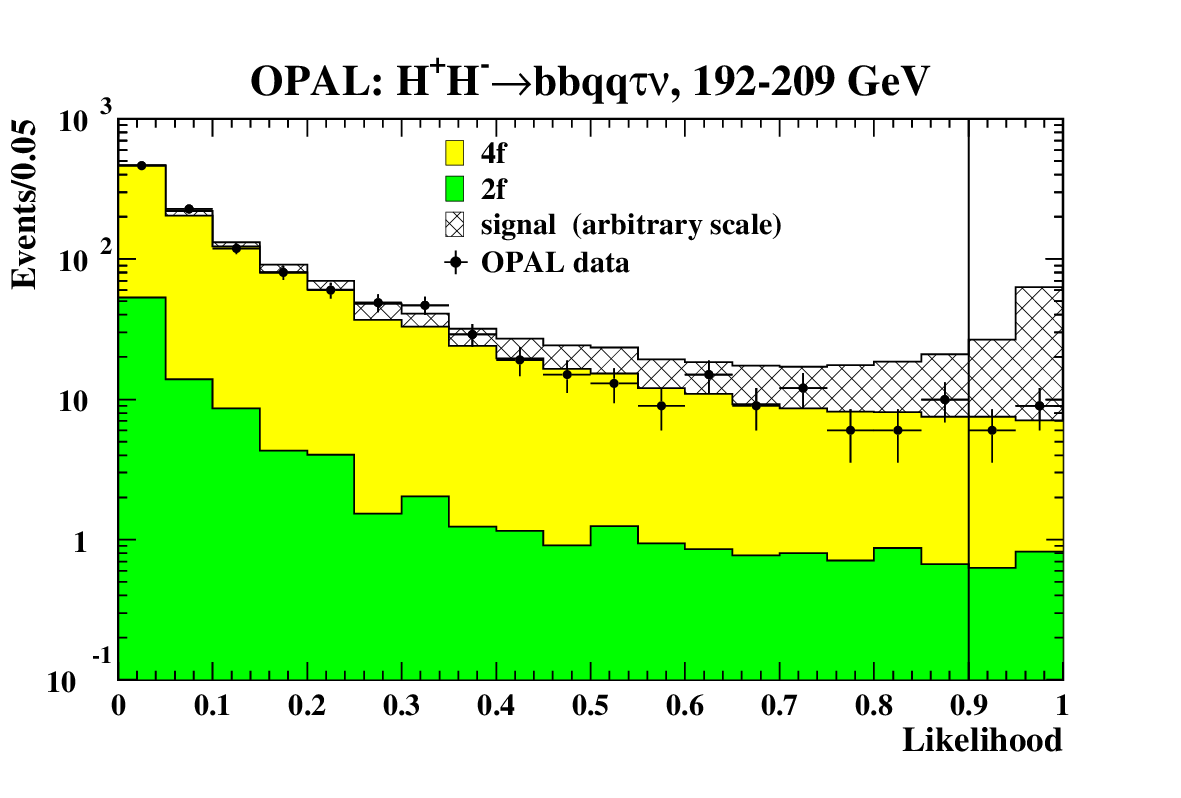, width=3 in}
\caption{\sl 
Likelihood output distribution for the \fourjt\ channel at (a) \sqrts=189~GeV
and (b) \sqrts=$192-209$~GeV. 
To form the signal histograms, the Monte Carlo distributions
are averaged for all simulated \mHmAc\ mass combinations in the mass range of
interest.
The expectations from  \SM\ processes are normalized to the
data luminosity. The lower cuts on the likelihood output are indicated by
vertical line. 
}

\vspace*{-7.2cm}

\hspace*{5.7cm} (a) \hspace*{6.85cm} (b) \vspace*{6.7cm}

\label{fig:watnlik}
\end{figure}

The likelihood output distributions are shown in
Figure~\ref{fig:watnlik}. 
There is an overall agreement between data and background distributions, apart
from a small discrepancy at 189 GeV.
Events are selected if their
likelihood output is larger than 0.9.  In total, 15 data events survive the
selection at $\sqrts=192-209$~GeV,  to be compared with 
$14.8\pm 0.6$~(stat.)~$\pm 1.9$~(syst.)  events expected from background sources. At \sqrts=189~GeV,
where the selection is optimized for low Higgs-boson masses, 13 data events are
selected with  $6.1\pm 0.5$~(stat.)~$\pm 1.3$~(syst.)  events expected.  The
contribution of four-fermion events, predominantly from semi-leptonic \WW\
production, amounts to 67\% at \sqrts=189~GeV and to 90\% at 
$\sqrts=192-209$~GeV.

At $\sqrts=192-209$~GeV, the signal selection efficiency starts at about 5\% at
\mHpm=40~GeV, reaches its maximum of about 40\% (depending on the mass
difference  $\Delta m=\mHpm-\mA$)  at \mHpm=60~GeV, then decreases to 12\% at
\mHpm=90~GeV. In the low-mass selection at \sqrts=189~GeV, the efficiency
depends strongly on the mass difference: at \mHpm=40~GeV, it is 
%% 22\% for $\Delta m=2$~GeV 
27\% for $\Delta m=2.5$~GeV and 60\% for  $\Delta m=10$~GeV. 
The selection efficiency
approaches its maximum at \mHpm=50~GeV 
%% (70\% for $\Delta m=10$~GeV) 
(73\% for $\Delta m=15$~GeV) and then
drops to zero at \mHpm=80~GeV. Table~\ref{tab:effi-WA} gives selection
efficiencies at representative \mHmAc\ points.

%%% rerunning the efficiency code in 2012 Jan:
%%% mH=40, mA=28: 61%
%%% mH=40, mA=30: 60%
%%% mH=40, mA=37.5: 27%
%%% mH=45, mA=42.5: 28%
%%% mH=50, mA=40: 69%
%%% mH=50, mA=35...36: 73%  (maximum)
%%% mH=55, mA=40: 71%

The systematic uncertainties 
due to the modeling of selection variables are evaluated 
with the method developed for the
\AWp\AWm\ channels and summarized in  Table ~\ref{tab:systau}.  

\begin{table}[htbp!]
\begin{center}
\begin{tabular}{|l||r|r|}\hline
Source & \multicolumn{2}{c|}{\fourjt} \\ 
\cline{2-3}
                               & signal    & background \\ 
\hline\hline
MC statistics                  & $\ge$2.7/$\ge$4.5  & 8.2/7.0 \\
\hline
preselection: tau ID           & 0.0       & 2.3/5.0 \\
\hfill other                   & 0.0/1.0 & 9.5/7.9 \\
\hline
likelihood selection: b-tag    & 0.3/1.4 & 2.4/3.0 \\
\hfill other                   & 3.2/2.1 & 18.4/11.2\\
\hline
\end{tabular}
\end{center}
\caption{\sl
Systematic uncertainties in percent for the \fourjt\ channel. Where two
values are given separated by a "/",  the first one belongs to the 189~GeV
selection and the second to the $192-209$~GeV selections. 
For the signal, the uncertainties due to the limited Monte Carlo statistics
are calculated by binomial statistics for a sample size of
500 events and they also depend, via the selection efficiency, on the assumed
Higgs-boson masses.
}
\label{tab:systau}
\end{table}
\begin{figure}[htbp!]
\centering
\vspace*{-12pt}
\epsfig{file=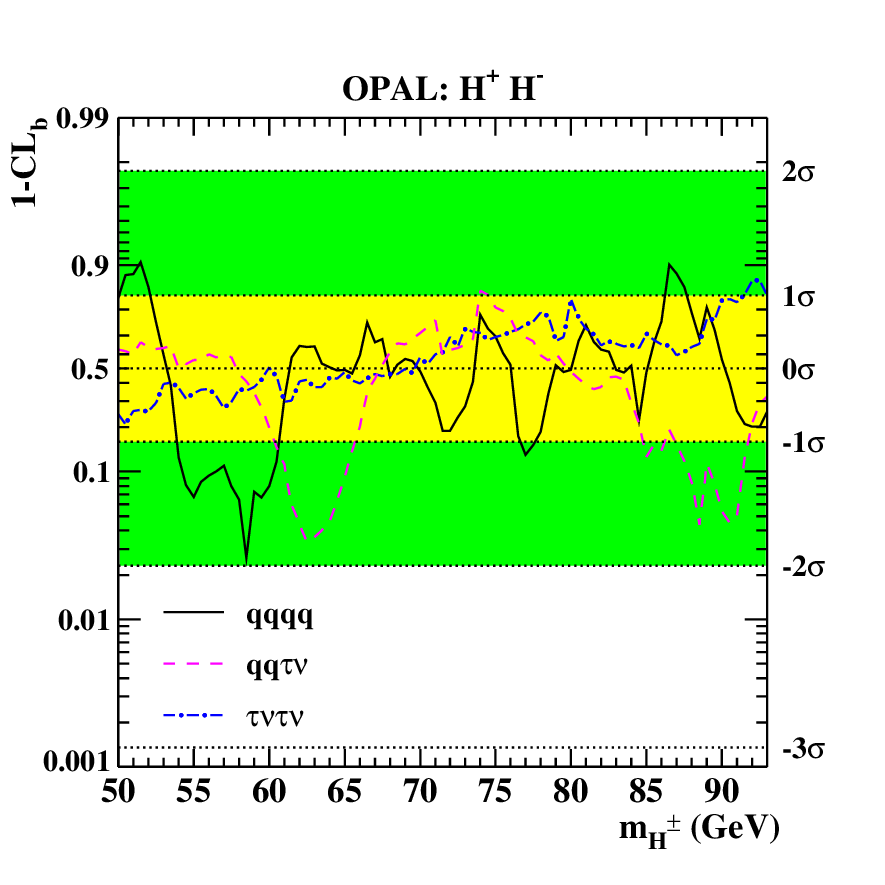, width=3. in}
\epsfig{file=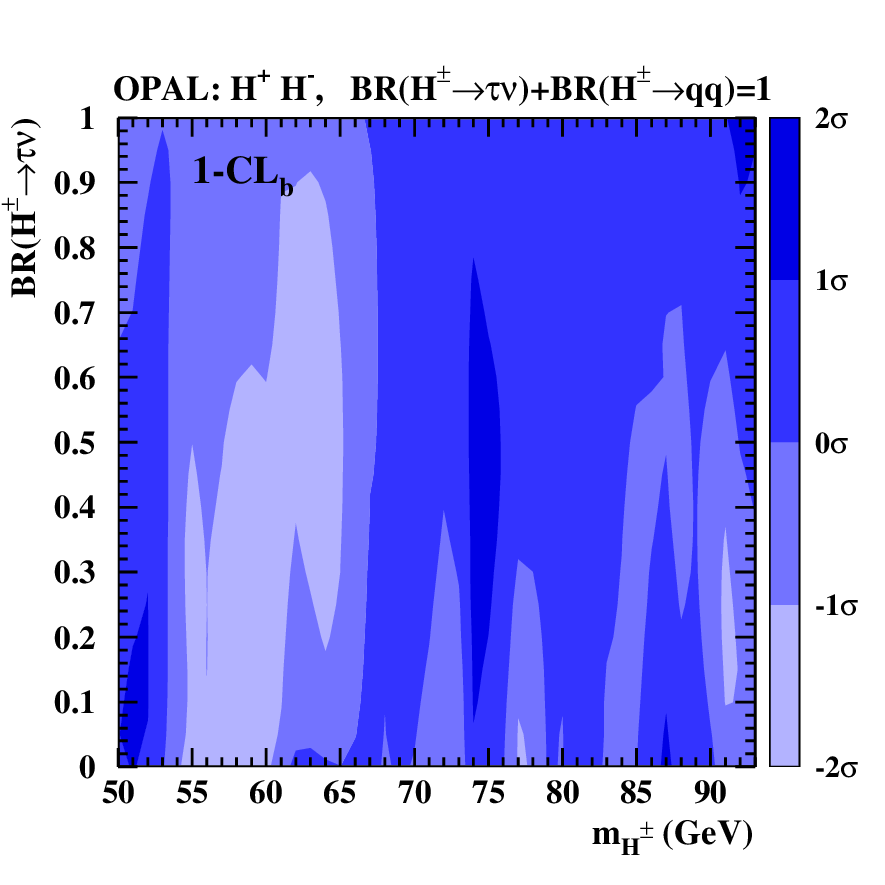, width=3. in}
\caption{\sl 
  The observed confidence levels for the background interpretation of the data, 
  $1-CL_b$, (a) for the three different final states  as a function of the
  charged Higgs-boson mass, and    (b) for the combined result in \THDMII\ 
  assuming \BRsum\ on the \mHBR\ plane. The significance values
  corresponding to the different shadings are shown by the bar at the right. 
  }

\vspace*{-4.7cm}

\hspace*{4.8cm} (a) \hspace*{6.85cm} (b) \vspace*{4.3cm}

\label{fig:clb}
\end{figure}
\section{Interpretation}\label{sect:interpret}

None of the searches has revealed a signal-like excess over the \SM\
expectation. The results presented here and those published 
previously~\cite{hpaper183,acoplanar2003} by the OPAL Collaboration are
combined using the method of~\cite{bock} to study the compatibility of the
observed  events with  ``background-only" and ``signal plus background"
hypotheses and to derive limits on charged Higgs-boson production.  The
statistical analysis is based on weighted event counting,  with the weights
computed from physical observables,  also called discriminating variables of
the candidate events (see Table~\ref{tab:discriminant}). 
Systematic uncertainties with
correlations are taken into account  in the confidence level ($p$-value)
calculations. To improve the sensitivity of the analysis, 
they are also incorporated into the weight definition~\cite{bock}
\footnote{For the weight definition, we use criteria (i) and (ii) in 
the 2HDM parameter scans and criterion (vii) in calculating model 
independent results. 
The generalized version of Eq. (2.9), given in Eq. (6.1), is used to include 
systematic errors in the event weights.
The treatment of correlations between systematic errors is discussed in section
5.1.}.

\begin{table}[htbp!]
\begin{center}
\begin{tabular}{|l|r||l|}
\hline
Channel & \sqrts\ (GeV) & Discriminant\\
\hline\hline
\twot & 183 & simple event counting\\
\twot & 189-209 & likelihood output\\
\twojt & 183-209 & reconstructed di-jet mass\\
\fourj & 183-209 & reconstructed di-jet mass\\
\hline
\eightj & 189 & simple event counting\\
\eightj & 192-209 & likelihood output\\
\sixjl & 189 & simple event counting\\
\sixjl & 192-209 & likelihood output\\
\hline
\fourjt & 189-209 & simple event counting\\
\hline
\end{tabular}
\end{center}
\caption{\sl Discriminating variables entering the statistical analysis 
for each search topology. Previously published results are also included.}
\label{tab:discriminant}
\end{table}

The results are interpreted in two different scenarios: in the traditional, 
super\-symmetry-favored \THDMII\ 
(assuming that there are no new additional light particles other than the Higgs
bosons) 
and in the \THDMI\ where under certain conditions fermionic
couplings are suppressed.

First, we calculate 
$1-CL_b$, the confidence~\cite{bock} under the background-only hypothesis, 
and then proceed to  calculate limits on the charged Higgs-boson  production
cross section in the signal + background hypothesis.  These results are used
to  provide exclusions in the model parameter space, and  in particular, on the
charged Higgs-boson mass.  

\subsection*{2HDM Type II}

First a general \THDMII\ is considered, where \BRsum. 
This model was thoroughly studied at LEP. 
It is realized in supersymmetric extensions of the \SM\ if no new
additional light particles other than the Higgs bosons are present. As our
previously published mass limit in such a model is
$\mHpm>59.5$~GeV~\cite{hpaper183}, only  charged Higgs-boson masses above
50~GeV are tested. Cross-section limits for lower masses can be found
in~\cite{hpmpaper172}. In this model, the results of the \twot, \twojt\ and
\fourj\ searches enter the statistical combination. 

The confidence $1-CL_b$ is plotted for each channel
separately in Figure~\ref{fig:clb}(a) and combined in Figure~\ref{fig:clb}(b). 
Note that $1-CL_b < 0.5$ translates to negative values of sigma 
(as indicated by
the dual y-axis scales in Figure~\ref{fig:clb}(a)) and indicates an excess of
events. 
%The largest deviation, a 2.8$\sigma$ excess observed
%in the  \twojt\ channel at \mHpm=88~GeV, comes from a known 
%deficiency~\cite{WW-OPAL2007} of
%the  Monte Carlo simulation of isolated tracks from the fragmentation and 
%hadronization process. When the three final states are
%combined assuming \BRsum, n
No deviation reaches the $2 \sigma$ level. 
\begin{figure}[htbp!]
\centering
\epsfig{file=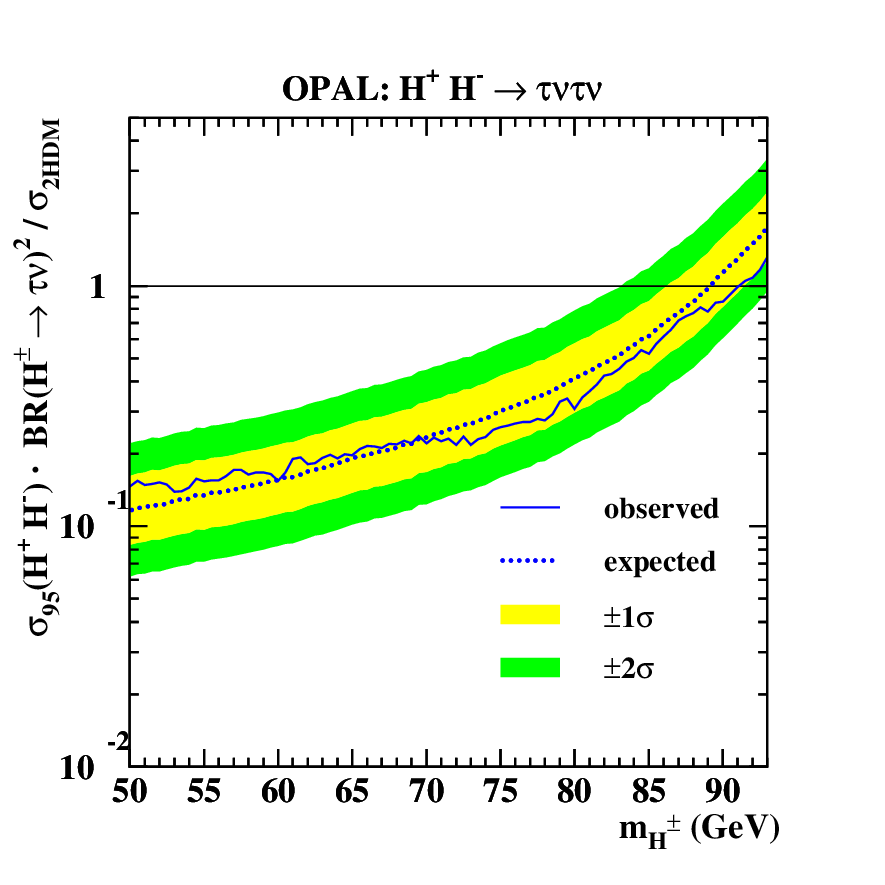, width=3. in}
\epsfig{file=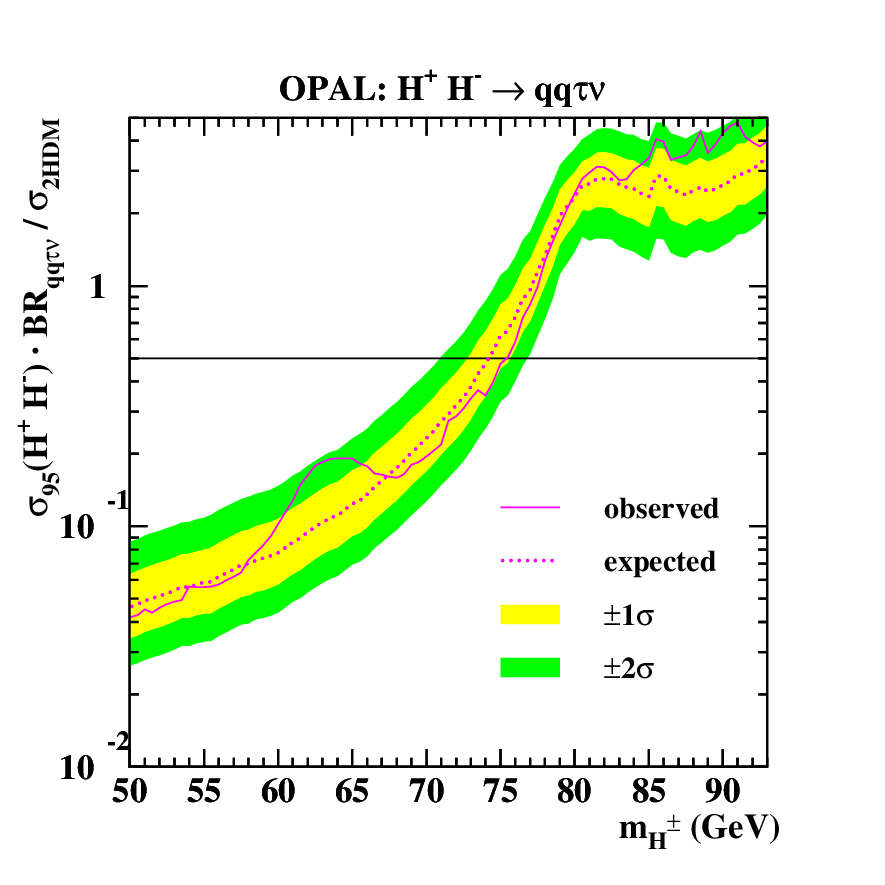, width=3. in}
\hspace*{0.5mm}\epsfig{file=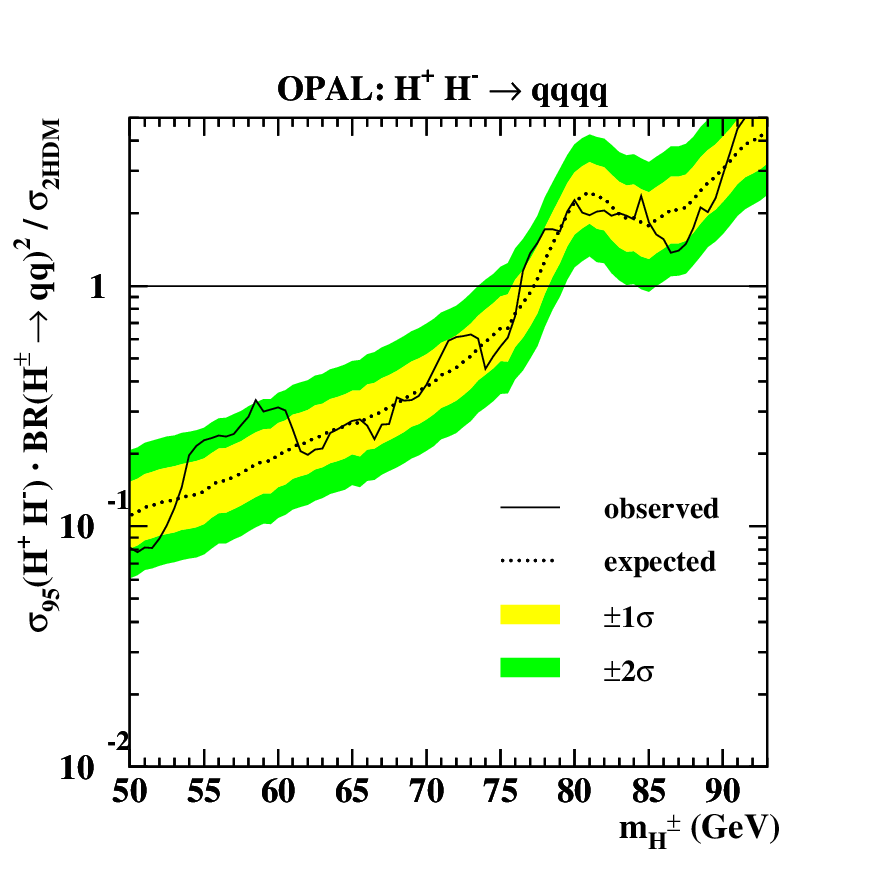, width=3. in}
\hspace*{1.5mm}\epsfig{file=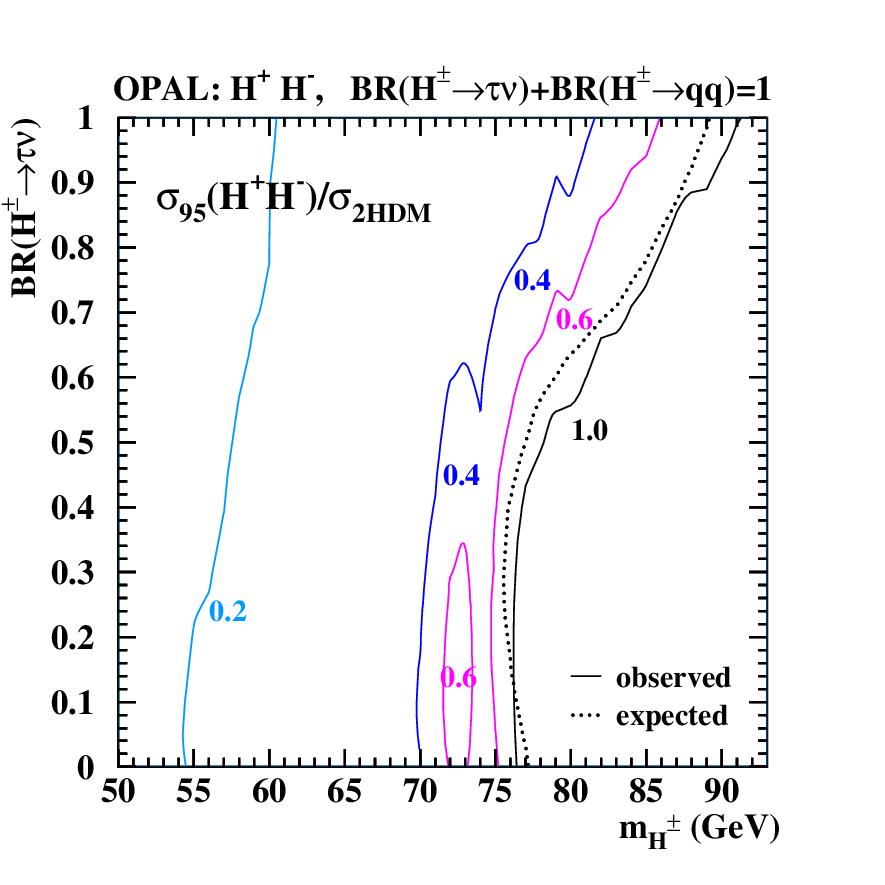, width=3. in}
\caption{\sl 
  Observed and expected  95\% \CL\ upper limits on the
  \HH\ production cross section times the relevant \Hpm\ decay branching
  ratios relative to the theoretical prediction 
  for the (a) \tnt\tnt\, (b) \qq\tnt\ and (c) \qq\qq\ channels.
  The horizontal lines indicate the maximum possible branching ratios for a
  given channel. In (b), $\mathrm{BR_{qq\tau\nu}} = 2\cdot\BRtn\cdot\BRqq$. 
  (d) Upper limits on the production cross section relative to the 2HDM
  prediction on the \mHBR\ plane in \THDMII\ assuming \BRsum. The plotted curves  
  are isolines along which the observed limit is equal to the number indicated.  
}

\vspace*{-13.2cm}

(a) \hspace*{7.0cm} (b) \hspace*{4.5cm} \vspace*{7.2cm}

(c) \hspace*{7.0cm} (d) \hspace*{4.5cm} \vspace*{5.5cm}

\label{fig:hphmxsec}
\end{figure}
\begin{figure}[htbp!]
\centering
\epsfig{file=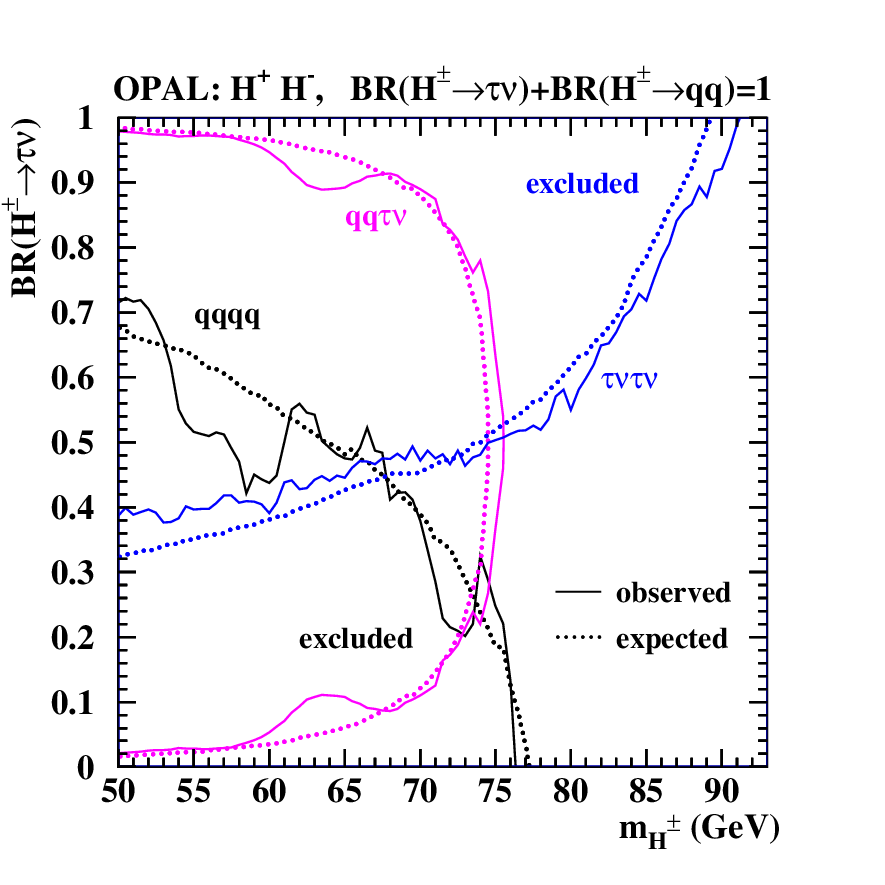,width=3. in}
\epsfig{file=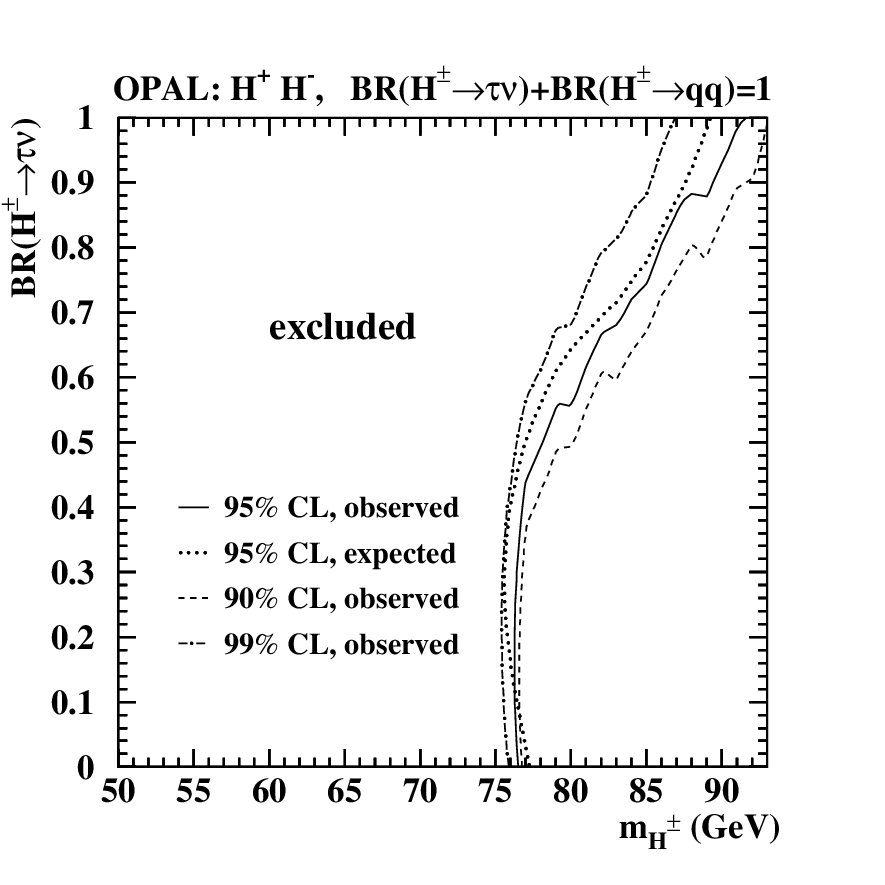,width=3. in}
\caption{\sl
  Observed and expected excluded areas at 95\% \CL\ on the
  \mHBR\ plane (a) for each search channel separately and 
  (b) combined in \THDMII\ assuming \BRsum. 
  For the combined result, the 90\% and 99\% \CL\
  observed limits are also shown. See Table~\ref{tab:masslim} for numerical
  values of the combined limit.
  }

\vspace*{-4.1cm}

\hspace*{4.6cm} (a) \hspace*{7cm} (b) \vspace*{4.5cm}

\label{fig:hphmres}
\end{figure}
\begin{table}[htbp!]
\begin{center}
\begin{tabular}{|c||c|c|}
\hline
      & \multicolumn{2}{c|}{Lower mass limit (GeV)} \\
      \cline{2-3}
\BRtn & Observed & Expected \\
\hline\hline
% -- submitted version
%0 & 76.9 & 77.9 \\
%0.5 & 79.2 & 78.0 \\
%0.65 & 82.0 & 81.7 \\
%1 & 91.2 & 89.2 \\
%\hline
%any & 76.6 (0.15) & 76.8 (0.27) \\
% -- after proper inclusion of systematic errors
0 & 76.5 & 77.2 \\
0.5 & 78.3 & 77.0 \\
0.65 & 81.9 & 80.5 \\
%0.653 & 82.0 & 80.7 \\
1 & 91.3 & 89.2 \\
\hline
any & 76.3 (0.15) & 75.6 (0.27) \\
\hline
\end{tabular}
\end{center}
\caption{\sl Observed and expected lower limits at 95\% \CL\
on the mass of the charged Higgs boson in \THDMII\ assuming \BRsum.
For the results independent of the branching ratio (last line), 
the \BRtn value at which the limit is set, is given in parenthesis.
}
\label{tab:masslim}
\end{table}

The results are used to set upper bounds on the charged Higgs-boson
pair production cross section relative to the 2HDM prediction as calculated
by HZHA. The limits obtained are shown for each channel 
separately in Figures~\ref{fig:hphmxsec}(a-c) 
and combined in Figure~\ref{fig:hphmxsec}(d).
The combined results are shown by ``isolines" 
along which $\sigma_{95}(\HH)/\sigma_\mathrm{2HDM}$, 
the ratio of the limit on the
production cross section and the 2HDM cross-section prediction,
is equal to the number indicated next to the curves.  

Excluded areas on the \mHBR\ plane  are presented 
for each channel separately in Figure~\ref{fig:hphmres}(a)
and combined in Figure~\ref{fig:hphmres}(b).  The expected mass limit from
simulated background experiments, assuming no signal, is also shown. For the 
combined results, the 90\% and 99\% \CL\ contours are also given. Charged
Higgs bosons are excluded up to a mass of 76.3~GeV at 95\% \CL, independent 
of \BRtn.
Lower mass limits for different values of \BRtn\
are presented in Table~\ref{tab:masslim}.

\subsection*{2HDM Type I}

We present here for the first time an interpretation of the OPAL charged 
Higgs-boson searches in an alternative theoretical scenario, a \THDMI. The novel
feature of this model with respect to the more frequently 
studied \THDMII\ is that the
fermionic decays of the charged Higgs boson can be suppressed. 
If the A boson is light, the \Hpm\ra\AWpm\ decay may play a crucial role. 

The
charged Higgs-boson sector in these models is described by three parameters:
\mHpm, \mA\ and \tanb. To test this scenario, the Higgs-boson decay branching
ratios \Hpm\ra\tnt, \cs, \cb, \AWpm\ and A\ra\bb\ 
are calculated by the program of Akeroyd et al.~\cite{THDM-Akeroyd},
and the model parameters are scanned in the range: 40~GeV~$\leq\mHpm\leq 94$~GeV, 
%12~GeV~$\leq\mA<\mHpm$,  $0.1\leq\tanb\leq 100$. Charged Higgs-boson
12~GeV~$\leq\mA<\mHpm$,  $0\leq\tanb\leq 100$. Charged Higgs-boson
pair production is excluded below 40~GeV by the measurement of the Z boson
width~\cite{Zwidth}. As the A boson detection is based on the identification of b-quark jets, no
%limits are derived for \mA$<2m_\mathrm{b}$.  Below \tanb=0.1, \BRAW\ vanishes
%and the limit is no longer sensitive to \mA.
limits are derived for \mA$<2m_\mathrm{b}$.

Both the fermionic (\twot, \twojt\ and \fourj) and the bosonic (\fourjt,
\sixjl\ and \eightj) final states play an important role and therefore their
results have to be combined. There is, however, a significant overlap between 
the events selected by the \HH\ra\qq\qq\ and
\HH\ra\AWp\AWm\ selections, and the events selected by the \HH\ra\qq\tnt\ and 
\HH\ra\AWpm\tnt\ selections.  Therefore, an automatic
procedure is implemented to switch off the less sensitive of the overlapping
channels, based on the calculation of the expected limit assuming no signal. In
general the fermionic channels are used close to the \mHmAc\ diagonal and for
low \tanb, and the searches for \Hpm\ra\AWpm\ are crucial for low values of
\mA\ and high values of \tanb.

%The confidence $1-CL_b$ is
%calculated combining the \sixjl\ and \eightj\ searches assuming \SM\
%branching ratios~\cite{PDG} for the \Wpms\ decay and is shown in
%Figure~\ref{fig:clb-wa}(a). The largest deviation 
%$1-CL_b=0.0009$ corresponding to $3.1\sigma$ is reached 
%at \mHpm=45 GeV and \mA=44.9~GeV (in the middle of the 
%very narrow lightest strip from \mHpm=40~GeV to 50~GeV 
%at the \mHmAc\ diagonal). 
%However, the mean background shift on the \mHmA\ plane amounts only
%to $1.1\sigma$. 
%$1-CL_b$ for the \fourjt\
%channel is shown in Figure~\ref{fig:clb-wa}(b).
The confidence $1-CL_b$ is calculated for the
combination of the \eightj\ and \sixjl\ searches and for the \fourjt\ search,
without requiring the \THDMI\ branching ratios 
(model independent scan), and for \tanb\ dependent combinations 
of all channels, including the fermionic ones, taking the \THDMI\
cross section predictions into account (model dependent scan).

\begin{figure}[htbp!]
\centering
\epsfig{file=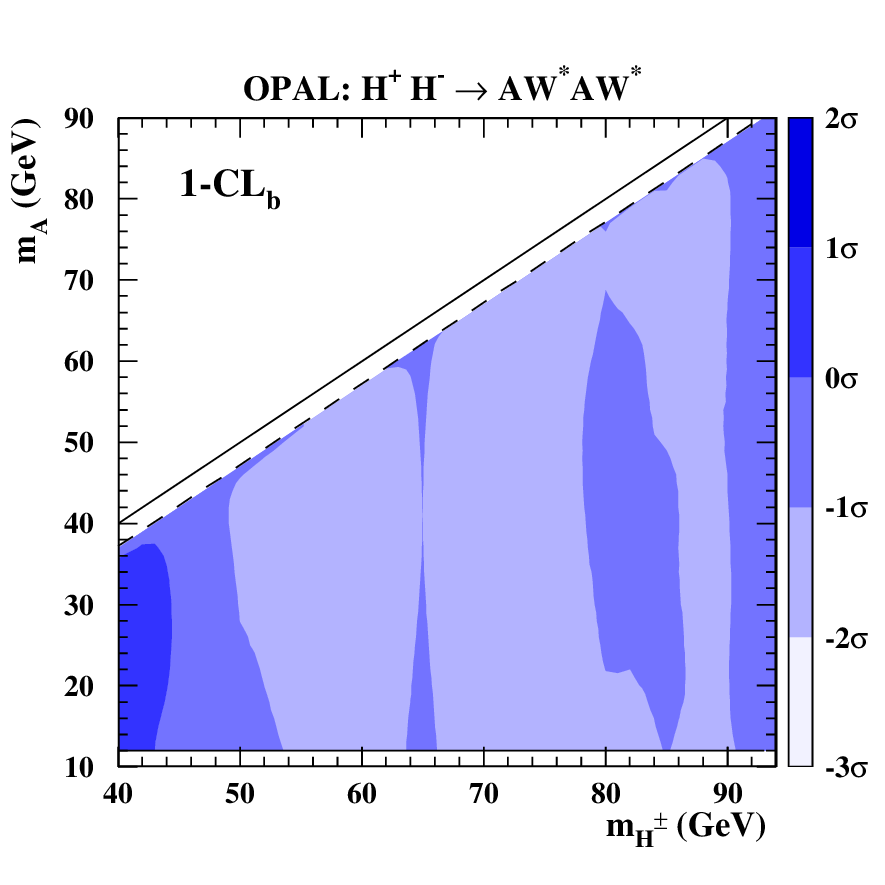, width=3. in}
\epsfig{file=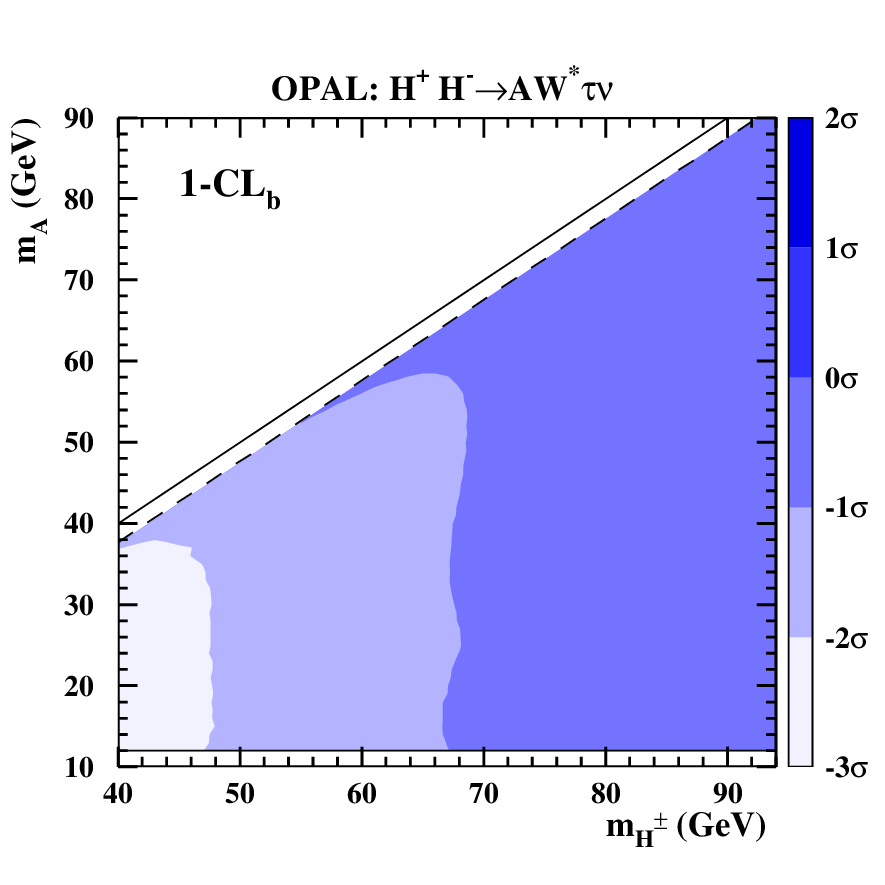, width=3. in}

\epsfig{file=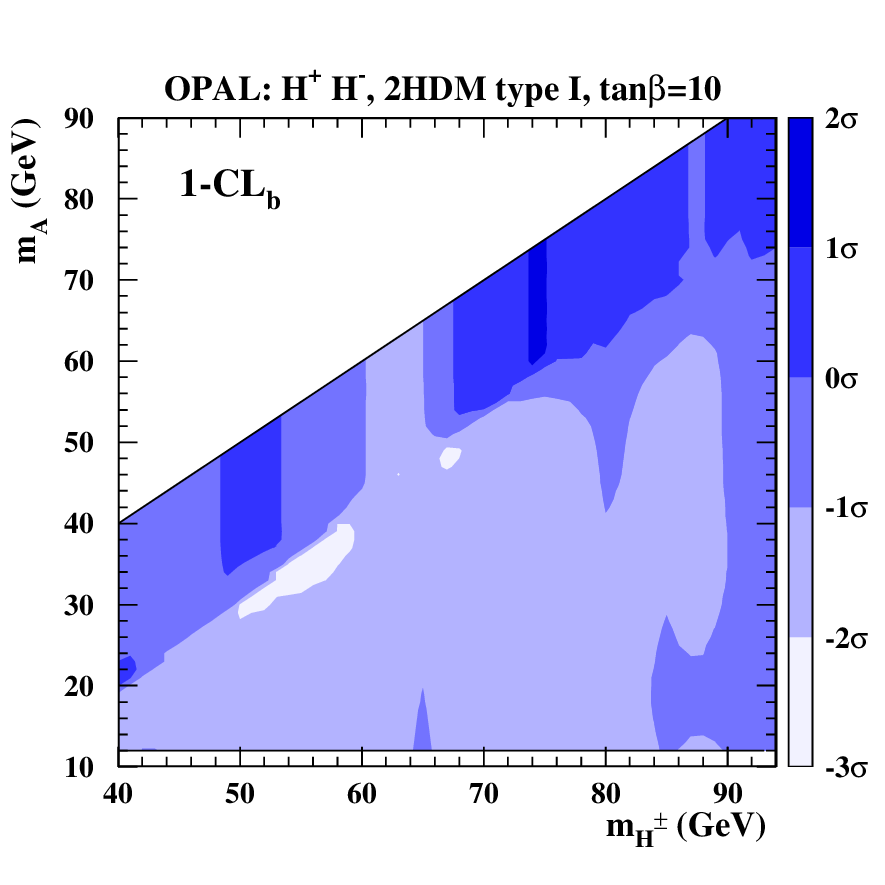, width=3. in}
\epsfig{file=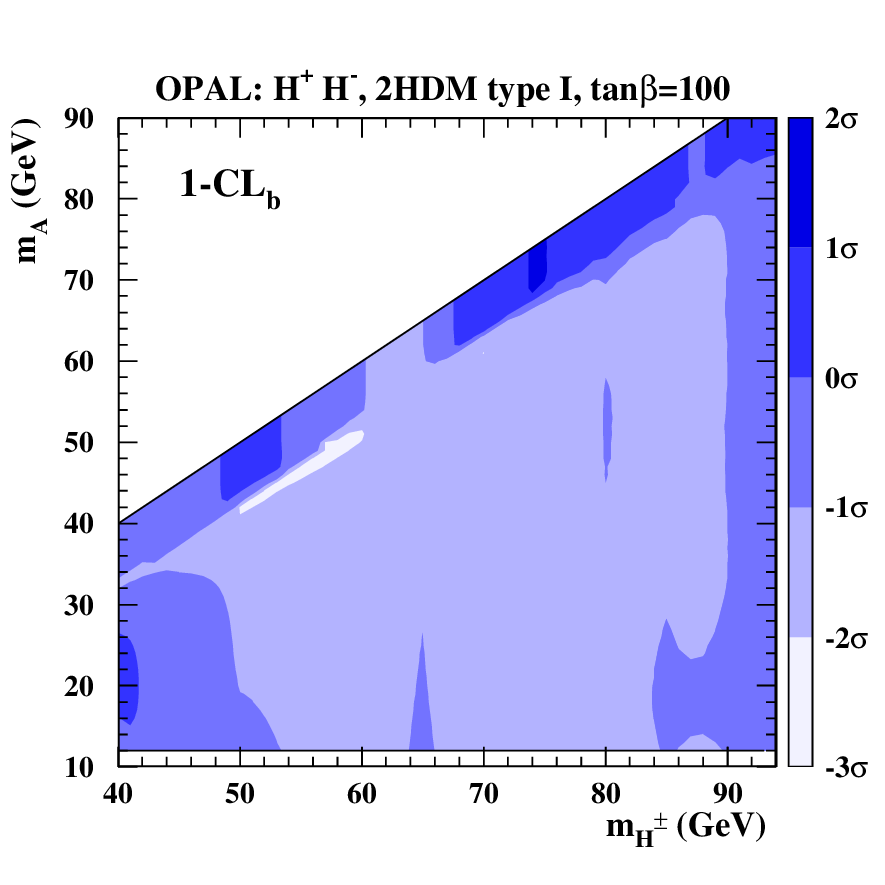, width=3. in}
\caption{\sl 
  The confidence $1-CL_b$  on the \mHmA\
  plane (a) for \HH\ra\AWp\AWm\
  combining the results of the \sixjl\ and \eightj\ searches, and (b)
  for \HH\ra\AWpm\tnt\ from the \fourjt\ search. 
  The combined results in \THDMI\ for (c) \tanb=10 and 
  (d) \tanb=100 are also shown.
  The significance values
  corresponding to the different shadings are shown by the bars at the right. 
}

\vspace*{-16.8cm}

\hspace*{4.6cm} (a) \hspace*{6.95cm} (b) \vspace*{7.2cm}

\hspace*{4.6cm} (c) \hspace*{6.95cm} (d) \vspace*{9.1cm}

\label{fig:clb-wa}
\end{figure}

The result for the \eightj\ and \sixjl\ combination is 
calculated assuming SM branching ratios~\cite{PDG} for the \Wpms\
decay and is shown in Figure~\ref{fig:clb-wa}(a).
Close to the \mHmAc\ diagonal, the eight- or 
six-jet structure of a \Hpm\ra\AWpm\ signal 
becomes less pronounced and the final state turns out four-jet-like 
with a few soft extra particles.
As the selection variables for the signal and the background become 
similar, the likelihood cut removes more signal events, resulting in a 
drop in efficiency and extrapolations towards the \mHpm=\mA\ limit 
are unreliable.
Morever, within the \THDMI, the branching ratio
for the bosonic Higgs decay vanishes at \mHpm=\mA.
Results for \mA$>$\mHpm$-3$~GeV are thus not included in 
Figure~\ref{fig:clb-wa}(a). In this mass region, the branching 
ratios for both \Hp\Hm\ra\eightj\ and 
\Hp\Hm\ra\sixjl\ are always less than $10^{-4}$. 
The largest deviation from background expectation, 
$1-CL_b= 0.029$, corresponding to $1.9 \sigma$ is reached at \mHpm=70~GeV
and \mA=64~GeV. Another local minimum at \mHpm=60~GeV and 
\mA=12~GeV, not visible in Figure~\ref{fig:clb-wa}(a), 
has $1-CL_b=0.052$. However, the mean background shift on the 
\mHmA\ plane amounts only to $1.1 \sigma$.
 
The $1-CL_b$ values for the \fourjt\ channel is shown in Figure~\ref{fig:clb-wa}(b).
Mass  combinations with \mA $>$ \mHpm$-2.5$~GeV 
are not included. In this mass region, the 2HDM(I) prediction for the 
H$^+$H$^- \rightarrow$ 4j+$\tau$ branching ratio is less 
than 0.005. 
The largest  deviation $1-CL_b=0.013$ corresponding to $2.2 \sigma$
appears for low charged Higgs-boson masses (\mHpm=40~GeV, \mA=21~GeV), 
reflecting the excess of events in the \sqrts=189~GeV search.
The mean background shift for this channel is $0.8 \sigma$.
%Note that the results shown in Figures~\ref{fig:clb-wa}(a-b) are
%model-independent.

%When all channels are combined within \THDMI, a few hot spots with a
%significance above $2\sigma$ survive. This is illustrated in
%Figures~\ref{fig:clb-wa}(c-d), 
%where the combined results are plotted for \tanb=10 and 100. 
%For \tanb=10, the largest excess $1-CL_b=0.014$ corresponding 
%to $2.2\sigma$ is found at \mHpm=55~GeV and \mA=34~GeV 
%(just before switching from the bosonic to the fermionic channels).
%This excess would correspond to a signal rate of 28.5\% 
%of the \THDMI\ expectation.
When all channels are combined within the \THDMI, the confidence 
levels shown in Figures~\ref{fig:clb-wa}(c-d) are obtained. Close to the 
\mHmAc\ diagonal, the results are determined by the analysis of
the fermionic channels. The \THDMI\ predicts 
BR(H$^{\pm} \rightarrow \tau \nu_{\tau}) \approx 0.65$ for the branching ratio,
depending only weakly on \mHpm. The upper parts of 
Figures~\ref{fig:clb-wa}(c-d) correspond thus to an almost horizontal cut 
in Figure 6(b) at \BRtn=0.65.
The lower parts of Figures 9(c-d) are essentially 
weighted combinations of the results in Figures 9(a-b),
depending on \tanb\ and the masses involved.
However, it has to be noted that also the decays \HH\ra\tpnu\tmnu\ are 
included and that the event weights in the statistical analysis 
are somewhat different for the model independent scans in 
Figures~\ref{fig:clb-wa}(a-b) 
and the model dependent scans in Figures~\ref{fig:clb-wa}(c-d).
In general, excesses in Figures~\ref{fig:clb-wa}(a-b) add up to excesses 
less than $2 \sigma$ in the combination.
A few regions with a significance above $2 \sigma$ are present.
For \tanb=10, the largest excess $1-CL_b=0.014$, corresponding
to $2.2 \sigma$, is found at \mHpm=55~GeV and 
\mA=34~GeV (just before switching from the bosonic to the
fermionic channels). This excess corresponds to a signal
rate of 28.5\% of the \THDMI\ expectation.  
Noting that the event weights depend on the hypothetical signal rate,
structures in the $1-Cl_b$ distribution as a function of 
$\Delta m = \mHpm - \mA$ are due to the 
similar increase of the signal cross-section with $\Delta m$ for different 
\mHpm\ values. The rise of the cross-section is steeper for larger \tanb, 
therefore the low $1-CL_b$ region shrinks from Figure~\ref{fig:clb-wa}(c) 
to Figure \ref{fig:clb-wa}(d).

In the limit of small \mA\ and large values of \tanb,
the Higgs decay into the \tnt\ channel is suppressed.
The structures of the $1-CL_b$ bands close to \mA=12~GeV 
in Figures ~\ref{fig:clb-wa}(a) and ~\ref{fig:clb-wa}(d) are therefore 
very similar.

As mentioned previously, the \Hpm\ra\AWpm\ decay becomes dominant if the A boson is 
sufficiently light. The smaller \tanb\ is, the smaller \mA\ should be. 
This is clearly seen from the structure of the result in
Figures~\ref{fig:clb-wa}(c-d): 
for \tanb=10, the bosonic decay becomes dominant at 
\mA\  $\lessapprox$  \mHpm$-$18~GeV, 
while for \tanb=100, it dominates already at 
\mA\  $\lessapprox$  \mHpm$-$6~GeV.

\begin{figure}[t!]
\centering
\vspace*{-12pt}
\epsfig{file=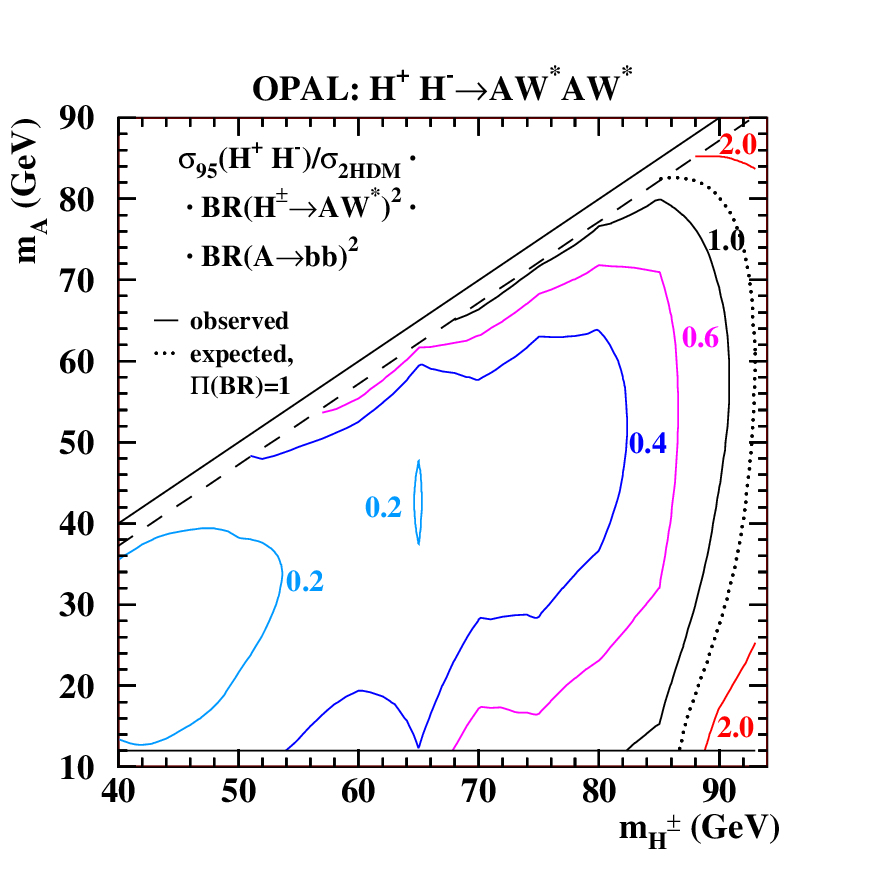, width=3. in}
\epsfig{file=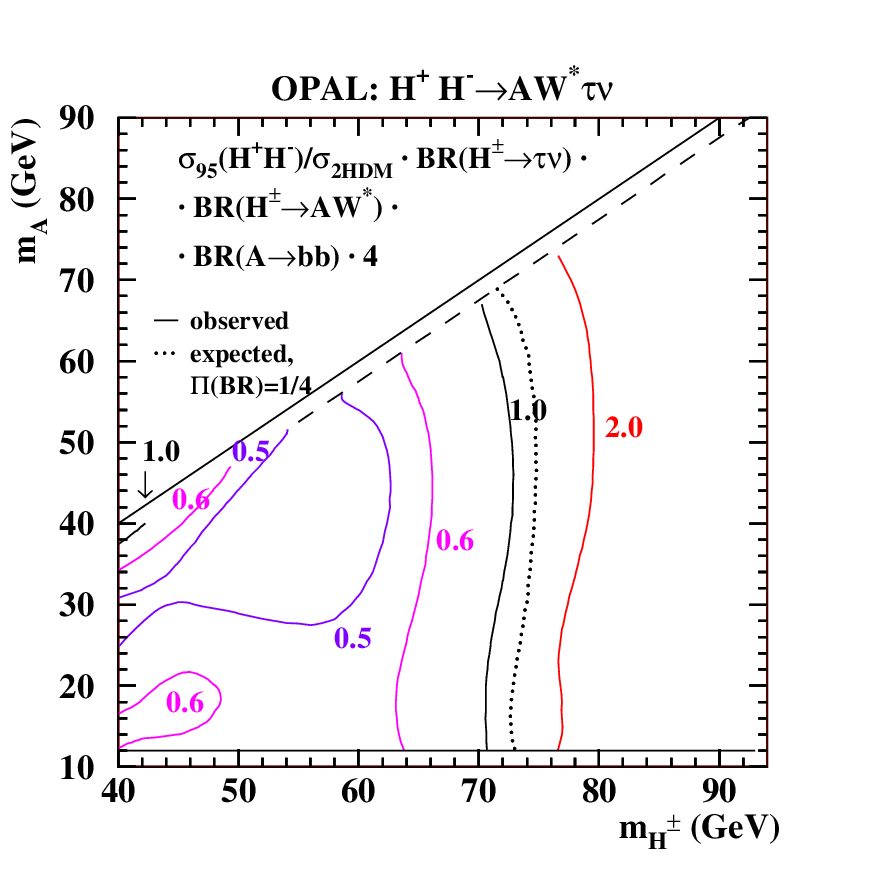, width=3. in}
\caption{\sl 
  The 95\% \CL\ upper limits on the production cross section 
  times relevant \Hpm\ and A boson decay branching ratios relative to the 
  2HDM prediction on the \mHmA\ plane for the process (a) \HH\ra\AWp\AWm\ and 
  (b) \HH\ra\AWpm\tnt.  
  The plotted curves are isolines along which the observed limit is 
  equal to the number indicated. 
  The expected limits are given for 
  (a) $\BRAW^2 \cdot \BRAbb^2 = 1$ and (b) $\BRtn \cdot \BRAW \cdot \BRAbb =
  0.25$ corresponding to the maximal value in \THDMI. Please note that the 
  plotted quantity on (b) is scaled by 4 to take into account this maximal 
  branching fraction.
}

\vspace*{-10.3cm}

(a) \hspace*{5cm} \vspace*{4.35cm}

\hspace*{12.5cm} (b) \vspace*{5cm}

\label{fig:xsec-wa}
\end{figure}

The model-independent limits on the charged Higgs-boson production
cross section relative to the 2HDM prediction 
are presented in Figure~\ref{fig:xsec-wa}(a) for the \HH\ra\AWp\AWm\ and 
in Figure~\ref{fig:xsec-wa}(b) for \HH\ra\AWpm\tnt\ searches, with
the only assumption that \Wpms\
%decays with \SM\ branching ratios. The results combining
decays with \SM\ branching ratios. The exclusion line for 40\% of the total
production cross section in Figure~\ref{fig:xsec-wa}(a) forms an island 
around \mHpm=60~GeV and \mA=12~GeV, corresponding to the minimum of 
$1-CL_b$ at this point.

\begin{figure}[p!]
\centering
\epsfig{file=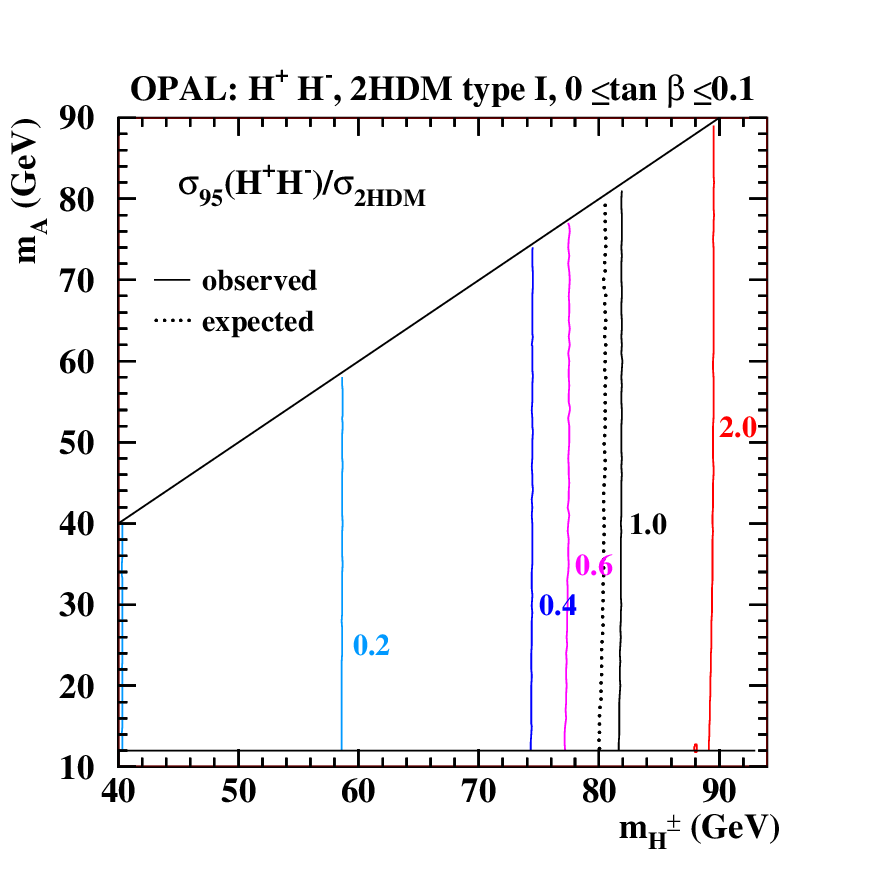, width=3. in}
\epsfig{file=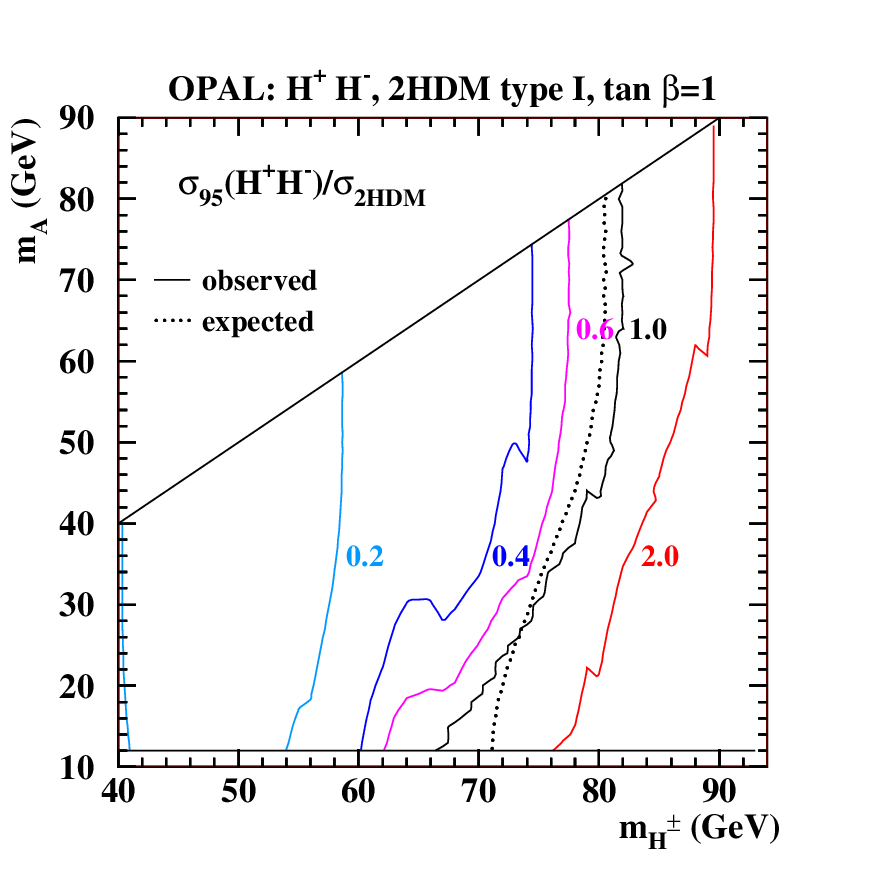, width=3. in}

\epsfig{file=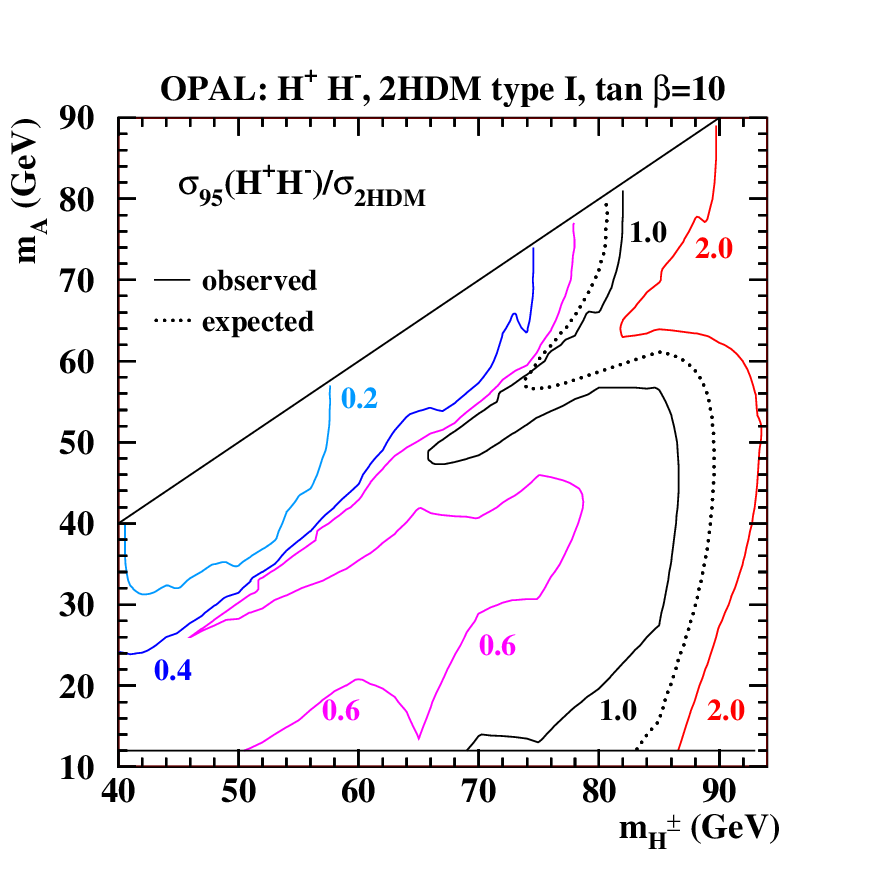, width=3. in}
\epsfig{file=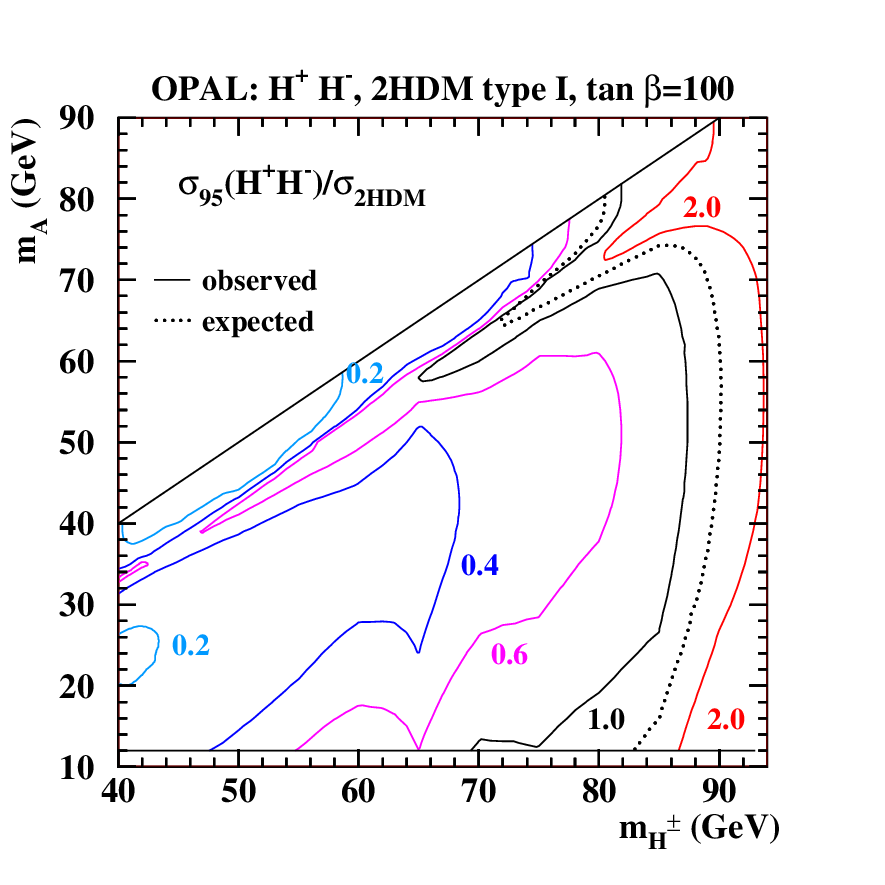, width=3. in}
\caption{\sl 
  The 95\% \CL\ upper limits on the production cross section  in
  \THDMI\ relative to the theoretical prediction on the \mHmA\ plane for
  different choices of \tanb: (a) 0.1, (b) 1.0, (c) 10.0 and (d) 100.0.
  The plotted curves are isolines along which the observed limit is 
  equal to the number indicated. 
}

\vspace*{-16.6cm}

\hspace*{4.9cm} (a) \hspace*{6.95cm} (b) \vspace*{7.2cm}

\hspace*{4.9cm} (c) \hspace*{6.95cm} (d) \vspace*{9.1cm}

\label{fig:xsec-2hdmI}
\end{figure}

The results combining
all channels using \THDMI\ branching ratios are shown in
Figure~\ref{fig:xsec-2hdmI} for different choices of \tanb. 
%
%For intermediate and large values of \tanb, an unexcluded region appears
%parallel to the \mHmA\ diagonal both in the observed and expected limits in
%Figure~\ref{fig:xsec-2hdmI} (see the full and dotted black curves corresponding 
%to a cross-section ratio of 1.0).   
%This can be understood by making two observations: Close to the diagonal, the
%eight- or six-jet structure of the \Hpm\ra\AWpm\ signal events becomes less 
%pronounced and the final state turns four-jet-like. As the selection variables
%for the signal and the background become similar, the likelihood cut removes 
%more signal events resulting in a drop in efficiency and therefore in sensitivity.
%This decreased sensitivity at the \mHmA\ diagonal is clearly visible in
%Figure~\ref{fig:xsec-wa}(a).
%On the other hand,  the rate of
%the fermionic events decreases in \THDMI\ by the distance from the 
%diagonal and by \tanb, so the fermionic searches also lose their sensitivity. 
For 0$\leq$\tanb$\leq$0.1, the excluded mass region is 
independent of \mA, since the AW$^{\pm *}$ final state 
does not contribute. The Higgs mass limit is identical to the
\THDMII\ limit at \BRtn=0.65.
This limit is also reached at the \mHmAc\ diagonal  
for any value of \tanb.
For intermediate and large values of \tanb, 
the boundary lines of the excluded mass regions have two local 
\mHpm\ minima.
% one at \mA=12~GeV and another one 
% at a region extending parallel to the \mHmAc\
% diagonal (see the full and dotted black curves corresponding
% to a cross-section ratio of 1.0).
% These minima reflect the loss of sensitivity for mass combinations
% without a pronounced \eightj\ or \sixjl\ structure and, in the
% second case, the decrease of the rate of fermionic events with 
% increasing mass difference \mHpm$-$\mA, at fixed \mHpm. 
% The sensitivity loss is also visible in Figure~\ref{fig:xsec-wa}(a).
The first minimum at \mA=12~GeV is due to the excess of events in the \AWp\AWm\
searches. The second minimum is at a region extending parallel to the \mHmAc\
diagonal (see the full and dotted black curves corresponding
to a cross-section ratio of 1.0). This reflects the loss of sensitivity in the 
\AWp\AWm\ searches for mass combinations without a pronounced \eightj\ or \sixjl\ structure 
and is also due to the channel switching procedure implemented to avoid the use of
overlapping events as explained above.

\begin{figure}[p!]
\centering
\epsfig{file=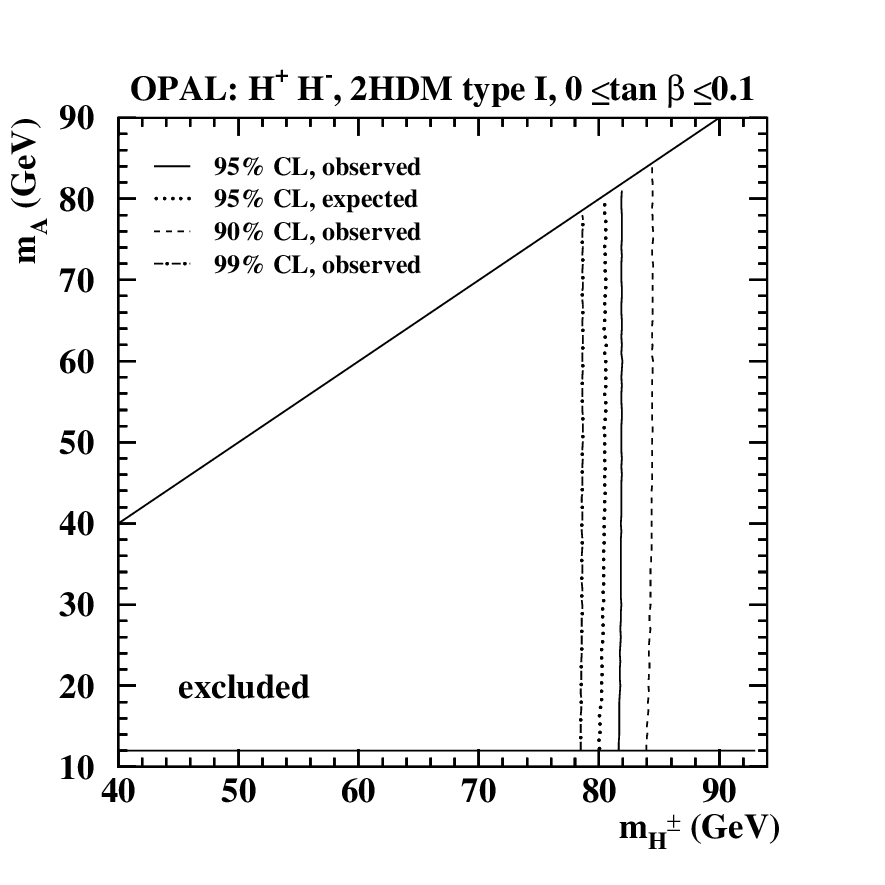, width=3. in}
\epsfig{file=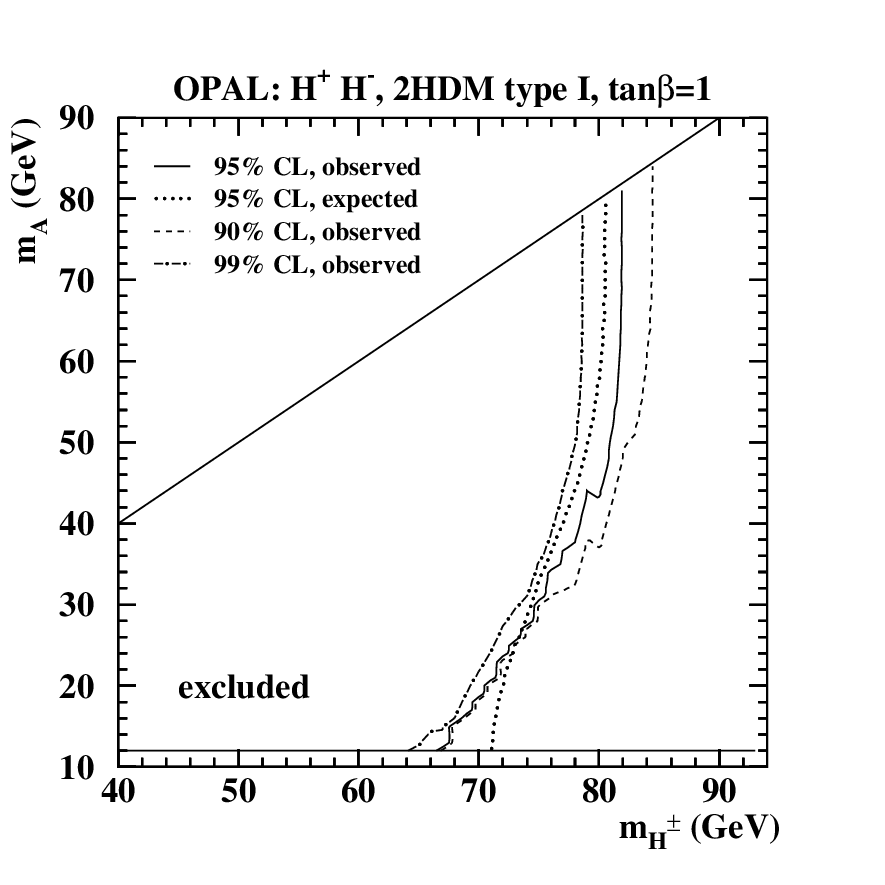, width=3. in}

\epsfig{file=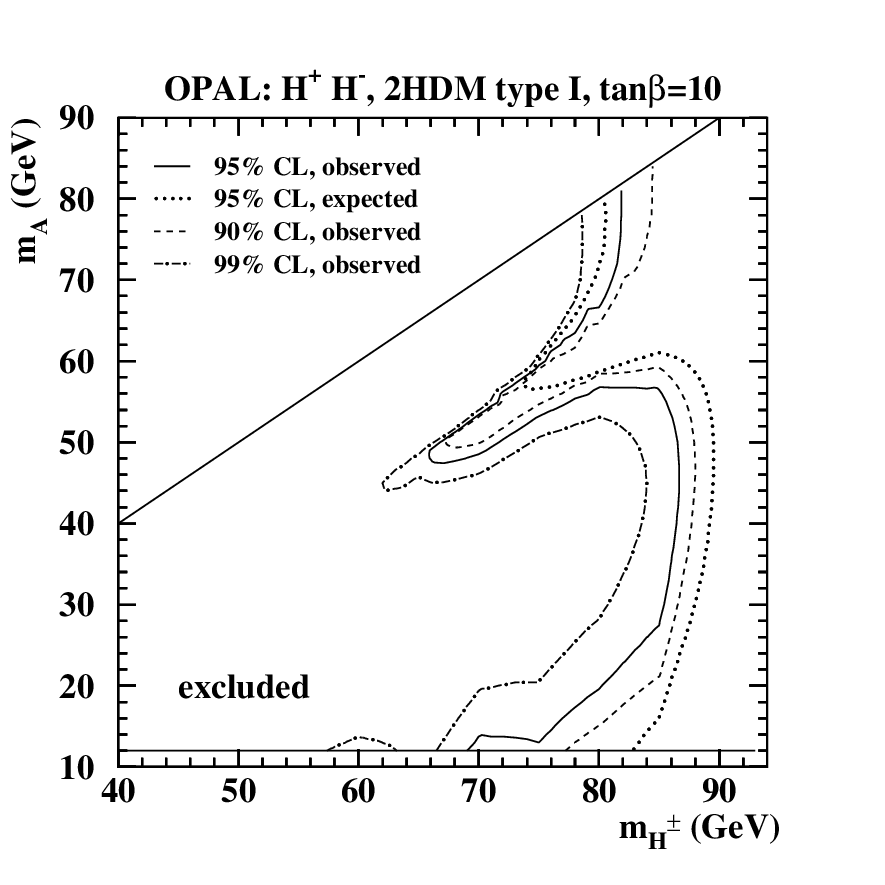, width=3. in}
\epsfig{file=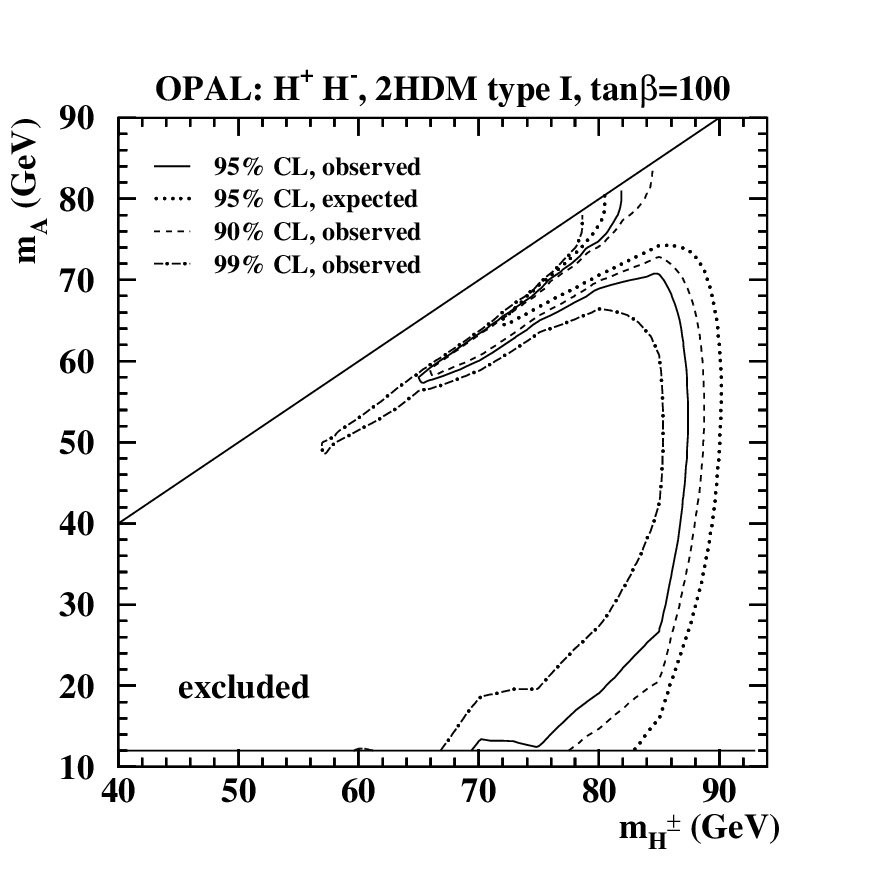, width=3. in}
\caption{\sl 
  Excluded areas at 90\%, 95\% and 99\% \CL\
  on the \mHmA\ plane in \THDMI\ for different choices of
  \tanb: (a) 0.1, (b) 1.0, (c) 10.0 and (d) 100.0.
}

\vspace*{-15.5cm}

\hspace*{4.9cm} (a) \hspace*{6.95cm} (b) \vspace*{7.2cm}

\hspace*{4.9cm} (c) \hspace*{6.95cm} (d) \vspace*{8.7cm}

\label{fig:excl-2hdmI}
\end{figure}

To further study the behavior of the unexcluded regions, 90\%, 95\% and 99\%
\CL\ excluded areas are shown in Figure~\ref{fig:excl-2hdmI} for different
%choices of \tanb.  At the 90\% \CL, the small unexcluded islands at
%\mA=12$-$15~GeV disappear and the observed limit (as expected) is set at the
%transition region discussed in the previous paragraph. 
choices of \tanb. The island at \mHpm=60~GeV can not be excluded 
at the 99\% \CL.

\begin{figure}[htbp!]
\centering
\epsfig{file=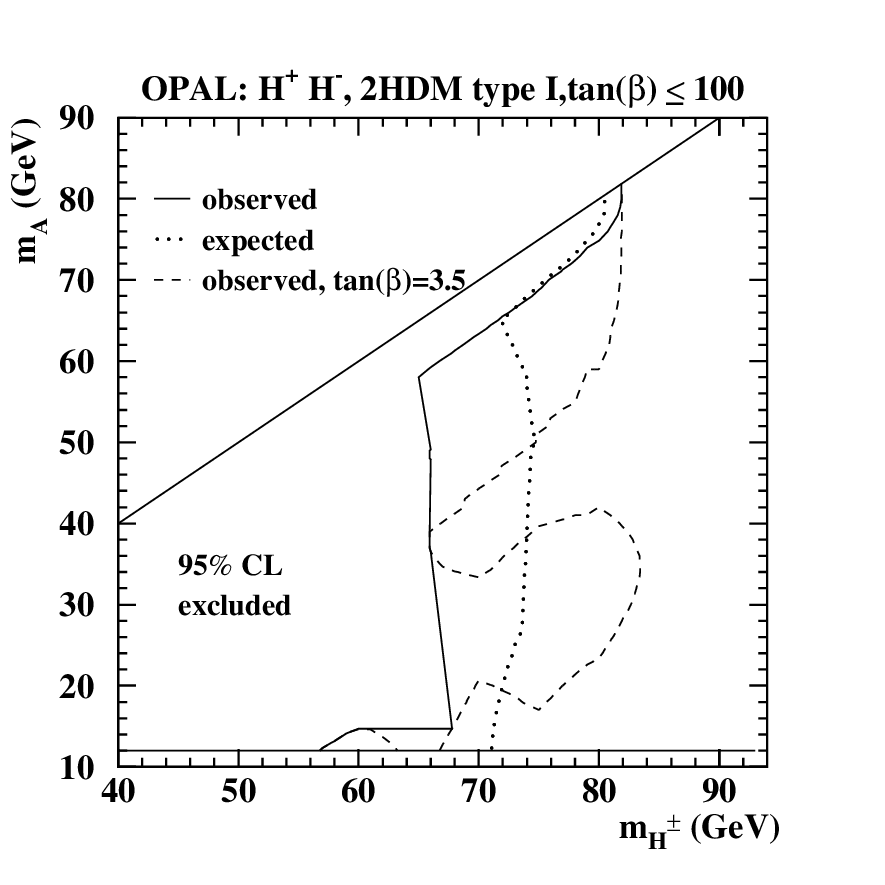, width=3. in} \hfill
\caption{\sl 
  Excluded areas in \THDMI\ on the \mHmA\ plane independent of \tanb\
  at 95\% \CL. The weakest overall mass limit is defined by the \tanb=3.5 
  exclusion, which is also shown. 
}
\label{fig:final-2hdmI}
\end{figure}

Due to the excess of events in the \HH\ra\AWp\AWm\ searches in the year 1999 data, 
the observed limit is lower than the expectation in all regions where the
\Hpm\ra\AWpm\ decay dominates.
%
%Finally our results are presented independent of \tanb\ in
%Figure~\ref{fig:final-2hdmI}, and the limits on the charged Higgs-boson mass are
%summarized in Table~\ref{tab:hwalimits}. The absolute lower limit on the
%charged Higgs-boson mass is 56.8~GeV for $\tanb\leq 100$ and 
%12~GeV~$\leq\mA\leq\mHpm$, which should be compared to an expectation of
%71.1~GeV. The observed limit is set by \tanb=3.5, where a small unexcluded 
%island is present around \mHpm$\approx$60~GeV and \mA$\approx$12~GeV 
%(also shown in  Figure~\ref{fig:final-2hdmI}).
%This unexcluded island is no longer present at 90\% \CL\ where
%the mass limit improves to 66.0~GeV.
Our final results are presented, for all \tanb, in 
Figure~\ref{fig:final-2hdmI}, and the limits on charged Higgs-boson mass 
are summarized in Table~\ref{tab:hwalimits}.
Since an excess is present in both the \eightj\ and \sixjl\ combination and 
the \fourjt\ channel, and the relative weighting of these channels 
depends on \tanb, the size of the mentioned island is
\tanb\ dependent. The absolute lower limit on the 
charged Higgs boson mass for 95\% \CL\ is set by \tanb=3.5, 
as indicated in Figure~\ref{fig:final-2hdmI}. It amounts to 56.8 GeV for 
$0\leq\tanb\leq100$ and 12~GeV$\leq$\mA$\leq$\mHpm, to be compared with an
expectation of 71.1~GeV. The unexcluded island is no longer present
at 90\% \CL\ where the observed mass limit improves to~66.0 GeV.

\begin{table}[htbp!]
\begin{center}
\begin{tabular}{|c|c||c|c|} 
\hline
  \tanb         & \mA & \multicolumn{2}{c|}{limit on \mHpm\ (GeV)}\\
  \cline{3-4}
                &     & observed & expected  \\   
\hline\hline
$\leq 100$ & 12~GeV$\leq\mA\leq\mHpm$      & 56.8 (3.5) &  71.1 (1.0)\\
                & \mA=12~GeV               & 56.8 (3.5) &  71.1 (1.0)\\
                & \mA=\mHpm/2              & 66.1 (3.5) &  73.9 (1.5)\\
                & $\mA\geq\mHpm-10$~GeV    & 65.0 (100) &  71.9 (100)\\
                & $\mA\geq\mHpm-5$~GeV     & 80.3 (100) &  77.3 (100)\\
\hline 
$\leq 0.1$      & 12~GeV$\leq\mA\leq\mHpm$       & 81.6  &  80.0 \\
                & \mA=12~GeV              & 81.6  &  80.0 \\
                & \mA=\mHpm/2             & 81.8  &  80.4 \\
                & $\mA\geq\mHpm-10$~GeV   & 81.9  &  80.5 \\
                & $\mA\geq\mHpm-5$~GeV    & 81.9  &  80.5 \\  
\hline
1        & 12~GeV$\leq\mA\leq\mHpm$        & 66.5  &  71.1 \\
                & \mA=12~GeV               & 66.5  &  71.1 \\
                & \mA=\mHpm/2              & 78.3  &  76.6 \\
                & $\mA\geq\mHpm-10$~GeV    & 81.9  &  80.5 \\
                & $\mA\geq\mHpm-5$~GeV     & 81.9  &  80.5 \\    
\hline
10       & 12~GeV$\leq\mA\leq\mHpm$        & 65.9  & 73.8 \\
                & \mA=12~GeV               & 69.0  & 82.8 \\
                & \mA=\mHpm/2              & 86.6  & 89.5 \\
                & $\mA\geq\mHpm-10$~GeV    & 81.3  & 79.4 \\
                & $\mA\geq\mHpm-5$~GeV     & 81.8  & 80.4 \\  
\hline
100      & 12~GeV$\leq\mA\leq\mHpm$        & 65.0  & 71.9 \\
                & \mA=12~GeV               & 69.4  & 82.9 \\
                & \mA=\mHpm/2              & 87.1  & 89.8 \\
                & $\mA\geq\mHpm-10$~GeV    & 65.0  & 71.9 \\
                & $\mA\geq\mHpm-5$~GeV     & 80.3  & 77.4 \\  
\hline
\end{tabular}
\caption{\sl Lower mass limits for the charged Higgs boson in
\THDMI. For the \tanb$\leq$100 results, 
the \tanb\ value at which the limit is set is indicated in parenthesis.
For any \tanb\ value, an extrapolation of the exclusion limits to \mHpm=\mA\
gives the result quoted in Table~\ref{tab:masslim} for \BRtn=0.65.}   
\label{tab:hwalimits}
\end{center}
\end{table}

For \mA$>$15~GeV, the \tanb-independent lower limit on the charged Higgs-boson mass
at 95\% \CL\ is 65.0~GeV with 71.3~GeV expected. 
The limit is
found in the transition region where the bosonic and fermionic channels have 
comparable sensitivities. The 6~GeV difference is due to the excess
observed in the \HH\ra\AWp\AWm\ search.
%On the \mHmAc\ diagonal, the \Hpm\ra\AWpm\ decay becomes kinematically
%suppressed and \BRtn$\approx 0.64-0.66$ depending on \mHpm. 
%Thus, for \mHpm=\mA, the \THDMII\ result for
%\BRtn=0.65 of 81.9~GeV (see Figure~\ref{fig:hphmres} and 
%Table~\ref{tab:masslim}) is reproduced. 

\section{Summary} 

A search is performed for the pair production of charged Higgs bosons in
electron-positron collisions at LEP2, considering the decays \Hpm\ra\tnt, \qq\
and \AWpm. No signal is observed. The results are interpreted in the framework
of Two-Higgs-Doublet Models. 

In \THDMII, required by the minimal supersymmetric extension
of the \SM,  charged Higgs bosons are excluded up to a mass of 
76.3~GeV (with an expected
%limit of 75.6~GeV) when \BRsum\ is assumed. See Figure~\ref{fig:hphmres} and
%Table~\ref{tab:masslim} for \BRtn-dependent limits. 
limit of 75.6~GeV) when \BRsum\ is assumed.
\BRtn-dependent limits are given in Figure~\ref{fig:hphmres} and 
Table~\ref{tab:masslim}.

In \THDMI,  where
fermionic decays can be suppressed and \Hpm\ra\AWpm\ can become dominant,  a
\tanb-independent lower mass limit of 56.8 GeV is observed for \mA$>12$~GeV (with
an expected limit of 71.1~GeV) due to an excess observed at
\sqrts=192$-$202 GeV in the \HH\ra\AWp\AWm\ search, discussed in
Section~\ref{sect:wawa}. 
For \mA$>15$~GeV, the observed limit improves to \mHpm$>65.0$ GeV 
(with an expected limit of 71.3 GeV). Figure~\ref{fig:final-2hdmI} shows the
excluded areas in the \mHmA\ plane and 
Table~\ref{tab:hwalimits} reports selected numerical results.

\input{pr427_acknowledge}
\input{pr427_ff_biblio}

\end{document}

%% file: pr427_f_defini.tex
\def\mrm{\mathrm}

\newcommand{\ra}{\mbox{$\rightarrow$}}
\newcommand{\sqrts}{\mbox{$\protect \sqrt{s}$}}
\newcommand{\pb}{\mbox{$\protect{\rm pb}^{-1}$}}
\newcommand{\etal}{\mbox{$et$ $al.$}}

\newcommand{\Hp}{\mbox{$\mathrm{H}^+$}}
\newcommand{\Hm}{\mbox{$\mathrm{H}^-$}}
\newcommand{\Hpm}{\mbox{$\mathrm{H}^{\pm}$}}
\newcommand{\Wpm}{\mbox{$\mathrm{W}^{\pm}$}}
\newcommand{\Wpms}{\mbox{$\mathrm{W}^{\pm *}$}}

\newcommand{\mA}{\mbox{$m_\mathrm{A}$}}
\newcommand{\mHpm}{\mbox{$m_{\mathrm{H}^{\pm}}$}}
\newcommand{\mWpm}{\mbox{$m_{\mathrm{W}^{\pm}}$}}
\newcommand{\tanb}{\mbox{$\tan\beta$}}

\newcommand {\HH}{\Hp\Hm}
\newcommand{\WW}{\mbox{$\mathrm{W}^{+}\mathrm{W}^{-}$}}
\newcommand{\ZZ}{\mbox{$\mathrm{ZZ}$}}
\newcommand{\ee}{\mbox{$\mathrm{e}^{+}\mathrm{e}^{-}$}}
\newcommand{\tpnu}{\mbox{${\tau^+\nu_{\tau}}$}}
\newcommand{\tmnu}{\mbox{${\tau^-{\bar{\nu}}_{\tau}}$}}
\newcommand{\tnt}{\mbox{${\tau\nu_{\tau}}$}}
\newcommand{\lnl}{\mbox{${\ell\nu_{\ell}}$}}
\newcommand{\lnu}{\mbox{${\ell\nu}$}}
\newcommand{\bb}{\mbox{$\mathrm{b} \bar{\mathrm{b}}$}}
\newcommand{\qq}{\mbox{$\protect {\rm q} \protect \bar{\rm q}$}}
\newcommand{\cs}{\mbox{$\protect {\rm c} \protect \bar{\rm s}$}}
\newcommand{\cb}{\mbox{$\protect {\rm c} \protect \bar{\rm b}$}}
\newcommand{\lplm}{\mbox{$\ell^+ \ell^-$}}
\newcommand{\AWp}{\mbox{$\mathrm{AW^{+*}}$}}
\newcommand{\AWm}{\mbox{$\mathrm{AW^{-*}}$}}
\newcommand{\AWpm}{\mbox{$\mathrm{AW^{\pm *}}$}}

\newcommand{\BRtn}{\mbox{$\mathrm{BR}(\Hpm\ra\tnt)$}}
\newcommand{\BRqq}{\mbox{$\mathrm{BR}(\Hpm\ra\qq)$}}
\newcommand{\BRsum}{\mbox{$\mathrm{\BRtn + \BRqq = 1}$}}
\newcommand{\BRAW}{\mbox{$\mathrm{BR}(\Hpm\ra\AWpm)$}}
\newcommand{\BRAbb}{\mbox{$\mathrm{BR}(A\ra\bb)$}}

\newcommand{\mHBR}{\mbox{$\mathrm{[\mHpm , \BRtn]}$}}
\newcommand{\mHmA}{\mbox{$\mathrm{[\mHpm , \mA]}$}}
\newcommand{\mHtanb}{\mbox{$\mathrm{[\mHpm , \tanb]}$}}

\newcommand{\mHmAc}{\mbox{$\mathrm{(\mHpm, \mA)}$}}

\newcommand{\twot}{\mbox{$\mathrm{2\tau}$}}
\newcommand{\twojt}{\mbox{$\mathrm{2j+\tau}$}}
\newcommand{\fourj}{\mbox{$\mathrm{4j}$}}
\newcommand{\fourjt}{\mbox{$\mathrm{4j+\tau}$}}
\newcommand{\sixjl}{\mbox{$\mathrm{6j+\ell}$}}
\newcommand{\eightj}{\mbox{$\mathrm{8j}$}}

\newcommand{\mtest}{\mbox{$m_\mathrm{test}$}}
\newcommand{\mref}{\mbox{$m_\mathrm{ref}$}}
\newcommand{\mlow}{\mbox{$m_\mathrm{low}$}}
\newcommand{\mhigh}{\mbox{$m_\mathrm{high}$}}
\newcommand{\LH}{\mbox{${\cal L}$}}

\newcommand{\SM} {SM}
\newcommand{\THDMI}{2HDM(I)}
\newcommand{\THDMII}{2HDM(II)}
\newcommand{\CL}{CL}

\newcommand{\bevt}{\mbox{${{\mathcal B}_\mathrm{evt}}$}}
\newcommand{\btagvar}{\bevt}

%% file: pr427_f_author.tex
\begin{center}{\Large        The OPAL Collaboration
}\end{center}\bigskip
\begin{center}{
%begin authorlist PLEASE DO NOT DELETE THIS COMMENT
G.\thinspace Abbiendi$^{  2}$,
C.\thinspace Ainsley$^{  5}$,
P.F.\thinspace {\AA}kesson$^{  7}$,
G.\thinspace Alexander$^{ 21}$,
G.\thinspace Anagnostou$^{  1}$,
K.J.\thinspace Anderson$^{  8}$,
S.\thinspace Asai$^{ 22}$,
D.\thinspace Axen$^{ 26}$,
I.\thinspace Bailey$^{ 25}$,
E.\thinspace Barberio$^{  7,   p}$
% ...
%end authorlist PLEASE DO NOT DELETE THIS COMMENT
}\end{center}\bigskip
\bigskip
%begin institutes
$^{  1}$School of Physics and Astronomy, University of Birmingham,
Birmingham B15 2TT, UK
\newline
$^{  2}$Dipartimento di Fisica dell' Universit\`a di Bologna and INFN,
I-40126 Bologna, Italy
\newline
% ...
$^{  5}$Cavendish Laboratory, Cambridge CB3 0HE, UK
\newline
% ...
$^{  7}$CERN, European Organisation for Nuclear Research,
CH-1211 Geneva 23, Switzerland
\newline
$^{  8}$Enrico Fermi Institute and Department of Physics,
University of Chicago, Chicago IL 60637, USA
\newline
% ...
$^{ 21}$Department of Physics and Astronomy, Tel Aviv University,
Tel Aviv 69978, Israel
\newline
$^{ 22}$International Centre for Elementary Particle Physics and
Department of Physics, University of Tokyo, Tokyo 113-0033, and
Kobe University, Kobe 657-8501, Japan
\newline
% ...
$^{ 25}$University of Victoria, Department of Physics, P O Box 3055,
Victoria BC V8W 3P6, Canada
\newline
$^{ 26}$University of British Columbia, Department of Physics,
Vancouver BC V6T 1Z1, Canada
\newline
% ...
$^{  p}$ now at The University of Melbourne, Victoria, Australia
\newline
% ...
%end notes

%% file: pr427_acknowledge.tex
\section*{Acknowledgements}

We particularly wish to thank the SL Division for the efficient operation
of the LEP accelerator at all energies
 and for their close cooperation with
our experimental group.  In addition to the support staff at our own
institutions we are pleased to acknowledge the  \\
Department of Energy, USA, \\
National Science Foundation, USA, \\
Particle Physics and Astronomy Research Council, UK, \\
Natural Sciences and Engineering Research Council, Canada, \\
Israel Science Foundation, administered by the Israel
Academy of Science and Humanities, \\
Benoziyo Center for High Energy Physics,\\
Japanese Ministry of Education, Culture, Sports, Science and
Technology (MEXT) and a grant under the MEXT International
Science Research Program,\\
Japanese Society for the Promotion of Science (JSPS),\\
German Israeli Bi-national Science Foundation (GIF), \\
Bundesministerium f\"ur Bildung und Forschung, Germany, \\
National Research Council of Canada, \\
Hungarian Foundation for Scientific Research, OTKA T-038240, 
and T-042864,\\
The NWO/NATO Fund for Scientific Research, the Netherlands.\\

%% file: pr427_ff_biblio.tex
%=====================================================================%